\documentclass[%
 reprint,
superscriptaddress,
 amsmath,amssymb,
 aps,
pra,
]{revtex4-2}

\usepackage{graphicx}
\usepackage{dcolumn}
\usepackage{bm}
\usepackage{nicefrac}
\usepackage{float}
\usepackage{amsmath}
\usepackage{booktabs}
\usepackage[dvipsnames]{xcolor}
\usepackage{hyperref}
\usepackage[capitalize]{cleveref}
\hypersetup{colorlinks=true,citecolor=blue,linkcolor=blue,urlcolor=blue}
\usepackage{comment}
\usepackage{tikz}
\usepackage{ulem}
\usepackage{subcaption}
\usepackage[dvipsnames]{xcolor}

\DeclareMathOperator{\arctanh}{arctanh}

\begin{document}

\preprint{APS/123-QED}

\title{Detection methods for optimal target reflectivity estimation with two-mode squeezed vacuum probes}

\author{Harel Radia}
 \email{harel.radia@mail.huji.ac.il}
\affiliation{
 Racah Institute of Physics, The Hebrew University of Jerusalem, Jerusalem 91904, Givat Ram, Israel
}
\author{Tuvia Gefen}
 \email{tuvia.gefen@mail.huji.ac.il}
\affiliation{
 Racah Institute of Physics, The Hebrew University of Jerusalem, Jerusalem 91904, Givat Ram, Israel
}
\author{Nadav Katz}
 \email{nadav.katz@mail.huji.ac.il}
\affiliation{
 Racah Institute of Physics, The Hebrew University of Jerusalem, Jerusalem 91904, Givat Ram, Israel
}
\affiliation{Qarakal Quantum Ltd., Derech Menachen Begin 144, Tel Aviv 6492102, Israel}

\date{\today}

\begin{abstract}
Target reflectivity estimation using a two-mode squeezed vacuum (TMSV) probe offers a theoretical advantage over classical schemes, but realizing this potential under the measurement constraints of microwave platforms remains a central challenge. In this work, we study the precision limits for target reflectivity estimation
across different energy and loss regimes, while accounting for realistic measurement restrictions. We characterize the optimal measurements and identify a transition in their structure: above a specific reflectivity threshold, a parametric amplifier receiver is optimal, whereas below it, the optimal observables are two-mode squeezing generators. We then study the performance of Gaussian measurements. When restricted to standard local homodyne detection, the TMSV probe is highly non-optimal.
However, we show that suitable non-local Gaussian measurements can closely approach the quantum Cramér-Rao bound at the large noise limit.
These results demonstrate that near-optimal quantum target reflectivity estimation is achievable in various relevant noisy regimes, even under the restriction of Gaussian measurements.
\end{abstract}
\maketitle

\section{Introduction}
Target reflectivity sensing, or equivalently the estimation of a Bosonic loss parameter, is a central problem in optical metrology. It arises in a wide range of applications, including quantum illumination (QI) \cite{R1_doi:10.1126/science.1160627,R16_PhysRevLett.114.080503,tan2008quantum,R21_PhysRevLett.98.160401,R22_PhysRevResearch.3.L042039,R23_PhysRevResearch.4.L012014,R24_PhysRevResearch.6.013034, nair2020fundamental,R25_PhysRevLett.118.070803,R26_PhysRevA.90.052308,R35_Park2023,R36_Jonsson_2022,R37_PhysRevResearch.3.013006, zhong2025relation}, sensitive radar systems \cite{R47_Agarwal_2024,R22_PhysRevResearch.3.L042039,R45_Huard_2023,R46_Norouzi2025,R52_Zeng_2024,tham2024quantum}, quantum noise sensing \cite{ng2016spectrum,tsang2023quantum,R65_shi2023ultimate, gardner2025stochastic, shi2025quantum}, Lindblad and loss estimation \cite{adesso2009optimal,nair2018quantum,gardner2025lindblad, gorecki2025interplay,gardner2026stochastic} and transmon qubits readout \cite{R61_Siddiqi_2018,R62_Franco_2024}. This problem has been studied extensively in the context of quantum illumination, where entangling the probe with an ancillary mode can lead to up to 3 dB gain in reflectivity estimation over unentangled strategies
\cite{R16_PhysRevLett.114.080503,tan2008quantum,R21_PhysRevLett.98.160401,R22_PhysRevResearch.3.L042039,R23_PhysRevResearch.4.L012014,R24_PhysRevResearch.6.013034,R25_PhysRevLett.118.070803,R26_PhysRevA.90.052308,R35_Park2023,R36_Jonsson_2022,R37_PhysRevResearch.3.013006, zhong2025relation}.
For this task, the two-mode squeezed vacuum (TMSV) state was identified as an optimal entangled state \cite{gong2021fundamental,R36_Jonsson_2022,bradshaw2021optimal,R25_PhysRevLett.118.070803}. This state can be efficiently generated with current capabilities using a parametric amplifier, and several detection schemes have been proposed to attain the corresponding quantum advantage \cite{guha2009gaussian, zhuang2017optimum,gong2021fundamental, R37_PhysRevResearch.3.013006, shi2023fulfilling, reichert2023quantum}.
Most previous works have focused on the quantum illumination regime of low reflectivity. More recently, the TMSV state was shown to be optimal for reflectivity estimation for any parameter regimes of the thermal-loss channel \cite{R36_Jonsson_2022,gardner2026stochastic}. This raises the question of which detection schemes are optimal for this sensing task given a TMSV input state and a general thermal-loss channel. Furthermore, in microwave platforms, detection is predominantly restricted to Gaussian measurements \cite{Nori}. It is therefore desirable to understand what precision limits can be achieved with different Gaussian measurements, and how closely such measurements can approach the ultimate quantum limits.

We address these questions in this work. We first analyze
the quantum precision limits for reflectivity estimation using a TMSV probe  in a general thermal-loss channel. We then study the performance of different detection schemes and identify optimal measurements that saturate the quantum limits. 

The symmetric logarithmic derivative (SLD) operator \cite{braunstein1994statistical,R55_Gao_2014} is employed to derive optimal measurement observables. This analysis reveals a transition point: parametric amplifier receivers
(followed by number-resolving measurements) 
are optimal above a certain threshold reflectivity.
Below the threshold these receivers are not necessarily optimal, and we demonstrate a gap between their performance and the quantum limit.  
At the transition point, the optimal measurement is a non-local homodyne readout (same as the double homodyne scheme of Ref. \cite{R37_PhysRevResearch.3.013006}), which leads us to further analyze the performance of Gaussian measurements.
While local homodyne measurement are highly suboptimal,
we demonstrate that non-local Gaussian measurements can closely approach the quantum limits across various parameter regimes.
Specifically, we study the performance of non-local homodyne readout and an adaptive heterodyne-homodyne strategy and find that they approach the quantum limit at complementary noisy regimes.

The paper is structured as follows: Sec.~\ref{sec:II} presents the formulation of the problem and relevant phase-space and entanglement properties. In Sec.~\ref{sec:III}, we analyze quantum precision limits for target reflectivity estimation,
and compare them to local homodyne measurements.
In  Sec.~\ref{sec:IV} optimal detection strategies are studied: Sec.~\ref{sec:IV.A} presents a derivation of optimal quantum measurements based on the SLD. In this section the transition in the SLD structure is identified and analyzed. Secs.~\ref{susec:homodyne_meas_strategy} and~\ref{sec:adaptive_heterodyne_homodyne} study Gaussian measurements that achieve nearly optimal precision. Finally, Sec.~\ref{sec:V} summarizes the results and discusses possible future directions.

\section{Formulation and Properties of the problem \label{sec:II}}

\subsection{Phase-Space Description}

We consider a reflectivity metrology model based on a TMSV probe. We denote the two modes of the TMSV as the \textit{idler} and \textit{signal}, with the canonical quadrature vector defined as $\vec{q} = [Q_i, P_i, Q_s, P_s]^T$ (where $\left[Q_{l},P_{m}\right]=i\delta_{l,m}$ and with the annihilation operator defined as $\hat{a} = \nicefrac{1}{\sqrt{2}}(Q+iP)$). The subscripts $i$ and $s$ denote the idler and signal modes, respectively, a convention maintained throughout the paper. The covariance matrix of the TMSV is given by:
\begin{align}
\begin{split}
    \Sigma_{TMSV} 
    &= 
    \frac{1}{2} 
    \begin{bmatrix}
        \cosh(2r) \hat{\mathbb{I}} &  \sinh(2r) \hat{\sigma}_z \\ \\ 
        \sinh(2r) \hat{\sigma}_z    &  \cosh(2r) \hat{\mathbb{I}}           
    \end{bmatrix}       
    \\&=
    \begin{bmatrix}
        (n_r+\nicefrac{1}{2}) \hat{\mathbb{I}} &  \sqrt{n_r^2+n_r}\ \hat{\sigma}_z \\ \\ 
        \sqrt{n_r^2+n_r}\ \hat{\sigma}_z    &  (n_r+\nicefrac{1}{2}) \hat{\mathbb{I}}
    \end{bmatrix}
    \label{eq:cov_tmsv}
   \end{split} 
\end{align}
where $r\in\mathbb{R}$ is the squeezing parameter, $n_r=\sinh^2 r$ is the mean photon number in each mode, $\hat{\sigma}_{z}$ is the Pauli-$Z$ matrix, and $\hat{\mathbb{I}}$ is the $2\times 2$ identity matrix.

This state can be prepared by applying a parametric amplifier unitary
$\hat{S}(r)\equiv \exp\left[-i r \left(Q_i P_s+Q_sP_i \right)\right]$ to the vacuum state. Under this operation, the vacuum covariance matrix $\Sigma_0=\frac{1}{2}\hat{\mathbb{I}}_{4}$ (where $\hat{\mathbb{I}}_{4}$ is the $4\times4$ identity matrix) transforms as:
\begin{equation}
    \Sigma_{0}\mapsto S(r)\Sigma_{0}S(r)^{t},
\end{equation}
where $S(r)$ is the corresponding symplectic matrix: 
\begin{equation}
    S(r) = 
    \begin{bmatrix}
        \cosh(r) \hat{\mathbb{I}} &  \sinh(r) \hat{\sigma}_z \\
        \sinh(r) \hat{\sigma}_z    &  \cosh(r) \hat{\mathbb{I}}
    \end{bmatrix}.
    \label{eq:Pamp}
\end{equation}
We assume that $n_r$ is fully controllable and serves as the sole parameter characterizing the probe. 

Of the two modes, the \textit{signal} is directed toward a target, where it interacts with thermal environmental noise. The background thermal mean photon number $n_\beta$ depends on the operating frequency $f$ and the environmental temperature $T$ according to
$n_\beta = \frac{1}{e^\beta - 1}$,
where $\beta = \frac{hf}{k_BT}$.
We model this interaction as a thermal-loss channel. This is a single-mode Gaussian channel, 
and thus defined by its action on the covariance matrix 
($\Sigma$)
and mean vector 
($\vec{\mu}$)
of a single mode Gaussian state:
\begin{equation}
\begin{split}
    \Sigma &\mapsto \eta\Sigma + (1-\eta)\left(n_\beta + \frac{1}{2}\right)\hat{\mathbb{I}}, \\
    \vec{\mu} &\mapsto \sqrt{\eta} \vec{\mu},
\end{split}
\end{equation}
where $\eta$ is the target's effective reflectivity.
In our model, $\eta$ serves as the sole parameter characterizing the target.
It governs the proportion of thermal noise introduced into the system via thermalization.

For two-mode states, such as the TMSV, the channel acts exclusively on the signal mode. Consequently, the mapping of the joint covariance matrix becomes:
\begin{equation}
\begin{aligned}
    \Sigma_{TMSV}\mapsto &
    \begin{bmatrix}
        \hat{\mathbb{I}} & 0 \\ 0 & \sqrt{\eta}\ \hat{\mathbb{I}}
    \end{bmatrix}
    \Sigma_{TMSV}
    \begin{bmatrix}
        \hat{\mathbb{I}} & 0 \\ 0 & \sqrt{\eta}\ \hat{\mathbb{I}}
    \end{bmatrix}
    \\
    &+ (1-\eta)\left(n_\beta+\frac{1}{2}\right)
    \begin{bmatrix}
        0 & 0 \\ 0 & \hat{\mathbb{I}}
    \end{bmatrix},
\end{aligned}
\label{eq:Bs}
\end{equation}
where $0$ in this Eq. denotes the $2\times 2$ zero matrix.
We assume that $n_\beta$ and $n_r$ are known, while the target reflectivity $\eta$ is the unknown parameter we wish to estimate. The limit $\eta = 1$ corresponds to a perfect reflector, where the signal is fully reflected from the target without admitting environmental noise. Conversely, $\eta = 0$ corresponds to the absence of a target, where the probe beam is completely lost.

The sensing protocol proceeds as follows: a TMSV state is prepared and its signal mode then goes through a thermal encoding channel (Eq.~(\ref{eq:Bs})).
The output state remains Gaussian and its covariance matrix is given by
\begin{equation}
\Sigma =
\begin{bmatrix}
    (n_r + \nicefrac{1}{2}) \hat{\mathbb{I}} & \sqrt{\eta\ (n_r^2 + n_r)} \hat{\sigma}_z \\
    \\
    \sqrt{\eta\ (n_r^2 + n_r)} \hat{\sigma}_z & (n_\beta + \nicefrac{1}{2} - \delta_n \eta) \hat{\mathbb{I}}
\end{bmatrix},
\label{eq:Cov}
\end{equation}
where
$\delta_n = n_\beta - n_r$
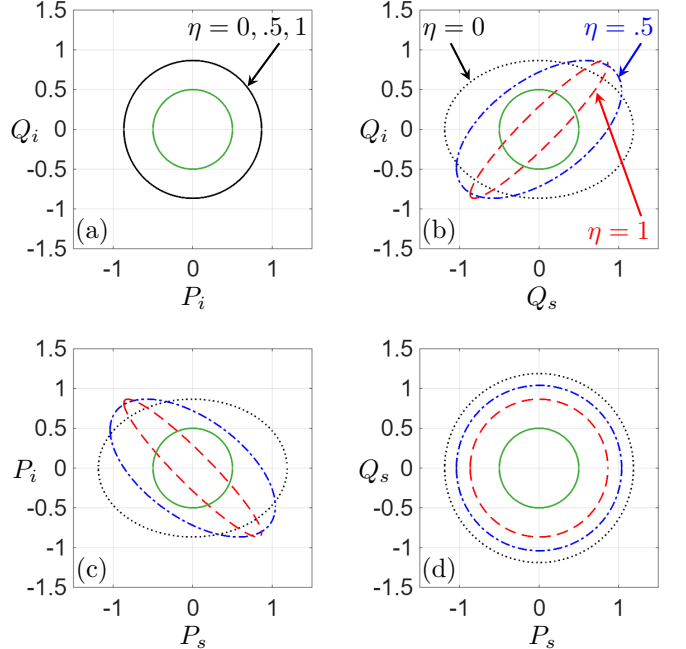
\begin{figure}[H]
    \centering
    \begin{tikzpicture}
        \node[anchor=south west, inner sep=0] (image) at (0,0) {
            \includegraphics[width=3.3in, trim={0mm 0mm 0mm 0mm}, clip]{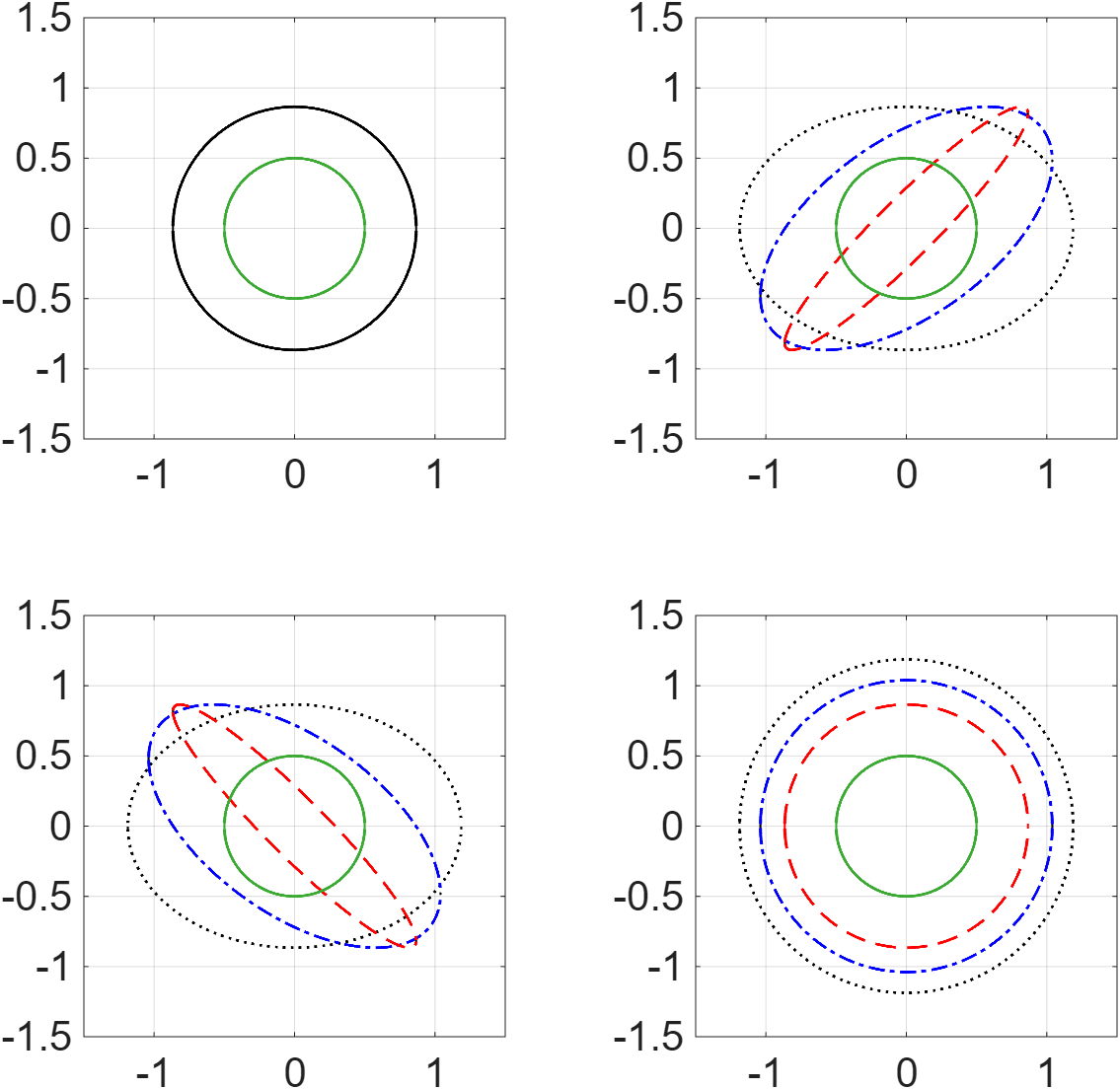}
        };

        \begin{scope}[x={(image.south east)}, y={(image.north west)}]

            \node[above,font=\normalsize,black] at (.105, .59) {(a)};
            \node[above,font=\normalsize,black] at (.652, .59) {(b)};
            \node[above,font=\normalsize,black] at (.105, .04) {(c)};
            \node[above,font=\normalsize,black] at (.652, .04) {(d)};
            
            \node[below,font=\normalsize] at (0, .82) {$Q_i$};
            \node[below,font=\normalsize] at (0.265, .55) {$P_i$};
            
            \node[above,font=\normalsize] at (.35, .92) {$\eta = 0,.5,1$};
            \draw[->, -stealth, thick] (.4, .93) -- (.35, .86);

            \node[below,font=\normalsize] at (0, .27) {$P_i$};
            \node[below,font=\normalsize] at (0.265, 0) {$P_s$};

            \node[below,font=\normalsize] at (0.55, .82) {$Q_i$};
            \node[below,font=\normalsize] at (0.815, .55) {$Q_s$};
            
            \node[above,font=\normalsize] at (.675, .92) {$\eta = 0$};
            \draw[->, -stealth, thick] (.67, .93) -- (.7, .87);
            
            \node[above,font=\normalsize,blue] at (.935, .92) {$\eta = .5$};
            \draw[->, -stealth, thick,blue] (.97, .93) -- (.935, .875);
            
            \node[above,font=\normalsize,red] at (.935, .59) {$\eta = 1$};
            \draw[->, -stealth, thick,red] (.97, .65) -- (.9, .85);

            \node[below,font=\normalsize] at (0.55, .27) {$Q_s$};
            \node[below,font=\normalsize] at (0.815, 0) {$P_s$};

        \end{scope}
    \end{tikzpicture}
    \caption{
        \label{fig:PS}  
        Phase-space projections of the output state for $n_\beta \approx 2.32$ ($T \approx 0.667\text{ K}$ at $5\ \text{GHz}$) and $n_r = 1$, across various target reflectivities $\eta$. The green line indicates the vacuum state for reference. The black, blue and red lines correspond to $\eta = 0, 0.5$ and $1$ respectively.
    }
\end{figure}

Fig.~\ref{fig:PS} shows phase-space projections of the output state onto various subspaces for different values of $\eta$, and illustrates how variation of $\eta$ modifies the state.
In this illustration, smaller $\eta$ leads to increased noise, as shown by the broadening of the signal mode (Fig.~\ref{fig:PS}(d)) and the cross quadrature plots (Fig.~\ref{fig:PS}(b,c)). Variation of $\eta$ also changes the orientation of the correlated quadratures, indicating a corresponding change in the normal modes. 

To better understand the effect of variation of $\eta$ on the state we use the Williamson decomposition, which identifies the independent normal modes and the effective occupation number of these modes.
According to the Williamson theorem ~\cite{williamson1936algebraic}, for any $2N\times2N$ covariance matrix $\Sigma$ there exists a real symplectic matrix, $M\in\text{Sp}\left(2N,\mathbb{R}\right)$, that diagonalizes it: $\Sigma=MDM^{t},$
where $D$ is a diagonal matrix $D=\text{diag}\left\{ \nu_{1},\nu_{1},...,\nu_{N},\nu_{N}\right\} $.
The diagonal elements $\left\{ \nu_{i}\right\}_{i=1}^{N}$
are the symplectic eigenvalues of $\Sigma$ and they represent the effective noise of the normal modes, where $\nu_{i}-1/2$ correspond to the thermal occupation number of the $i$-th normal mode.
In our case, the two symplectic eigenvalues are
\begin{equation}
\begin{aligned}
\nu_{\pm}
= \frac{1}{2}\bigg[
&\sqrt{
\delta_n^{2}\eta^{2}
-2\eta\left(n_{\beta}+n_{\beta}^{2}+n_r+n_r^{2}\right)
+\left(1+n_{\beta}+n_r\right)^{2}
} \\
&\pm (1-\eta)\delta_n
\bigg].
\end{aligned}
\label{eq:symp_eigenvalues}
\end{equation}
The diagonalizing symplectic matrix $M$ is given by the 
parametric amplifier transformation (Eq.~\ref{eq:Pamp})

with squeezing parameter $l$: 
\begin{equation}
    l=\frac{1}{2}\text{arctanh}\left(2\frac{\sqrt{\eta \left(n_{r}^2+n_{r}\right)}}{\left(1+n_{\beta}+n_{r}\right)+\eta\left(n_{r}-n_{\beta}\right)}\right).
 \label{eq:symp_transformation}   
\end{equation}

It can be thus seen that $\eta$ changes both the symplectic eigenvalues, i.e. the noise (illustrated by the width in Fig.~\ref{fig:PS}), and the symplectic transformation $M$ (the orientation of the cross-quadrature plots in Fig.~\ref{fig:PS}).
We show in Sec.~\ref{sec:review_FI} that the information about $\eta$
can be decomposed into information from the symplectic eigenvalues and information from the symplectic transformation, which in turn affects the structure of the optimal detection scheme.

\subsection{Entanglement Criterion}
For completeness, we review the parameter regimes in which the probe state remains entangled \cite{ferraro2005gaussian}. A two-mode Gaussian state is entangled if the smallest positive eigenvalue, $\varepsilon^-$, of the symplectic partial time-reversal matrix $\Sigma_{\text{sptr}}$ satisfies:
\begin{equation}
     \varepsilon^- < \frac{1}{2}
    \label{eq:eps}   
\end{equation}
Following Ref.~\cite{R60_Lloyd_2012} (Eq.~59), the symplectic partial time-reversal matrix is defined as:
\begin{equation}
    \Sigma_{\text{sptr}} = \Omega O_{\text{ptr}}^\dagger \Sigma O_{\text{ptr}}
\end{equation}
Here, the symplectic form $\Omega$ is given by:
\begin{equation}
\Omega =
\begin{bmatrix}
    -\hat{\sigma}_y & 0 \\
    0 & -\hat{\sigma}_y
\end{bmatrix}
\label{eq:omega}
\end{equation}
where $\hat{\sigma}_y$ is the Pauli-$Y$ matrix, and the partial time-reversal operator $\hat{O}_{\text{ptr}}$ is defined as
$\hat{O}_{\text{ptr}} = \mathrm{diag}\left(
    \begin{bmatrix}
        1 & 1 & 1 & -1
    \end{bmatrix}
    \right).$
Applying this to our model, the general condition in Eq.~(\ref{eq:eps}) reduces to the inequality:
\begin{equation}
    \Delta n = n_\beta - n_\eta < 0, \; \text{with } n_\eta = \frac{\eta}{1 - \eta}. 
    \label{eq:Dn}
\end{equation}
We will see that $n_\eta$ defined here is a key variable that appears throughout the paper in different contexts.

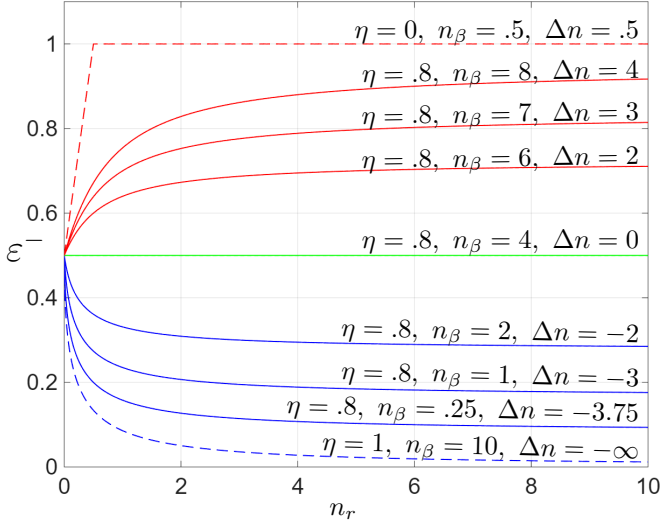
\begin{figure}[H]
    \centering
    \begin{tikzpicture}
        \node[anchor=south west, inner sep=0] (image) at (0,0) {
            \includegraphics[width=3.3in, trim={0mm 0mm 0mm 0mm}, clip]{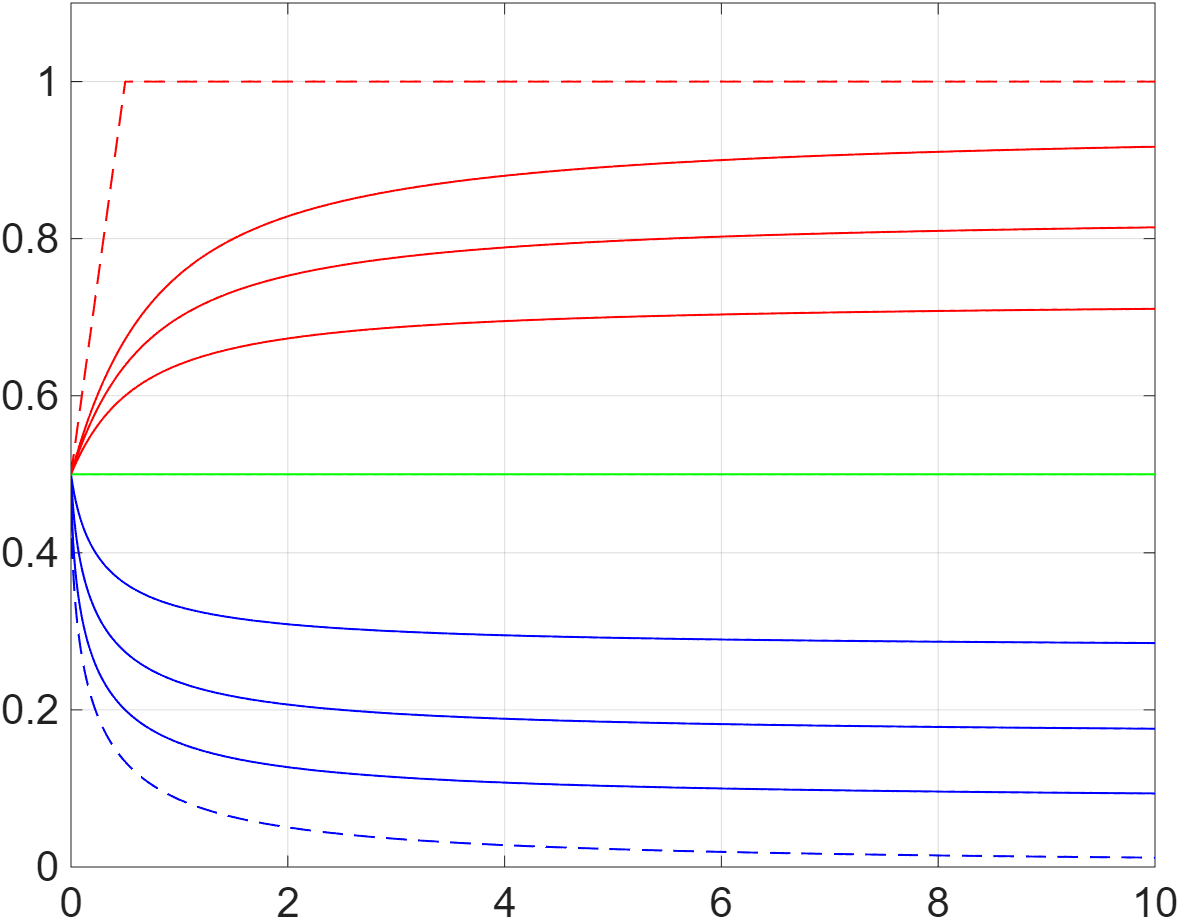}
        };

        \begin{scope}[x={(image.south east)}, y={(image.north west)}]

            \node[below,font=\Large] at (0, .55) {$\varepsilon^-$};
            \node[below,font=\normalsize] at (0.5, 0) {$n_r$};

            \node[left,font=\normalsize] at (.98, .935) 
            {$\eta = 0,\  n_\beta=.5,\ \Delta n =.5$};
            
            \node[left,font=\normalsize,rotate=1] at (.98, .855) 
            {$\eta = .8,\  n_\beta=8,\ \Delta n =4$};
            
            \node[left,font=\normalsize,rotate=.9] at (.98, .77) 
            {$\eta = .8,\  n_\beta=7,\ \Delta n =3$};   
            
            \node[left,font=\normalsize,rotate=.8] at (.98, .68) 
            {$\eta = .8,\  n_\beta=6,\ \Delta n =2$};    

            \node[left,font=\normalsize] at (.98, .51) 
            {$\eta = .8,\  n_\beta=4,\ \Delta n =0$};    

            \node[left,font=\normalsize,rotate=-1.2] at (.98, .32) 
            {$\eta = .8,\  n_\beta=2,\ \Delta n =-2$}; 

            \node[left,font=\normalsize,rotate=-1.4] at (.98, .23) 
            {$\eta = .8,\  n_\beta=1,\ \Delta n =-3$}; 

            \node[left,font=\normalsize,rotate=-1.3] at (.98, .16) 
            {$\eta = .8,\  n_\beta=.25,\ \Delta n =-3.75$}; 

            \node[left,font=\normalsize,rotate=-1.5] at (.98, .085) 
            {$\eta = 1,\  n_\beta=10,\ \Delta n =-\infty$};

        \end{scope}
    \end{tikzpicture}
    \caption{
    \label{fig:ent_cond}
    Dependence of the smallest symplectic eigenvalue $\varepsilon^-$ on the squeezing parameter $n_r$, plotted for various values of the photon number difference $\Delta n$. Solid blue curves indicate the entangled regime ($\Delta n < 0$). Solid red curves correspond to the separable regime ($\Delta n > 0$). For reference, the dashed curves depict the limiting cases of perfect reflection ($\eta \to 1$, blue) and total loss ($\eta \to 0$, red).
    }
\end{figure}

To demonstrate the relationship between squeezing and entanglement, we analyze the behavior of $\varepsilon^-(n_r)$ at $\eta = 0.8$ ($n_\eta = 4$) across varying $n_\beta$. As depicted by the solid blue curves in Fig.~\ref{fig:ent_cond}, the system successfully enters the entangled regime when $\Delta n < 0$. Here, the lowest symplectic eigenvalue acts as a continuous measure of non-classical correlations, directly mapped to the logarithmic negativity $E \equiv \max[0, -\ln(2\varepsilon^-)]$ \cite{Adesso}; as $\varepsilon^-$ drops further below $0.5$, the degree of entanglement increases monotonically with the applied squeezing. 
In the limit of perfect reflectivity ($\eta \to 1$, $n_\eta \to \infty$), represented by the dashed blue curve, $\varepsilon^- \to 0$ as $n_r$ increases. This implies that the non-zero asymptotic (Appendix~\ref{apx:eps}) limit observed in the general case (solid blue curves) arises solely from thermalization noise.

Conversely, when $\Delta n > 0$ (solid red curves), the system remains entirely separable. Because the state never crosses the threshold ($\varepsilon^- \ge 0.5$), the logarithmic negativity evaluates identically to zero and ceases to be a relevant metric. This demonstrates that if the necessary environmental criteria are not met, entanglement cannot be achieved 
even for arbitrarily large squeezing. At the opposite extreme, where $\eta \to 0$ (dashed red curve), the probe beam is completely lost to the environment. This total loss yields $\varepsilon^- > 0.5$ for all $n_r > 0$, confirming that entanglement cannot be sustained in this limit. The derivation and further analysis of $\varepsilon^-$ is provided in Appendix~\ref{apx:eps}.

\section{Precision limits for reflectivity estimation\label{sec:III}}
 We focus on precision limits for estimating $\eta.$ Using the Cramér-Rao bound (CRB)~\cite{kay1993statistical}, we quantify the precision limits with the Fisher information (FI) and quantum Fisher information (QFI)~\cite{braunstein1994statistical}. Let us first review the theory of Cramér-Rao bounds and apply them to our problem.

\subsection{Review of classical and quantum Cramér-Rao bounds}
\label{sec:review_FI}
Consider a probability distribution $p(x|\theta)$ that depends on a parameter $\theta$. The classical CRB states that the mean squared error $\sigma_\theta^2$ of any unbiased estimator for $\theta$ is lower-bounded by the inverse of the FI, denoted as $I^{-1}$: $\sigma_\theta^2 \geq I^{-1}$, where the FI is defined as~\cite{kay1993statistical}:
\begin{equation}
I=\int\frac{\left(\partial_{\theta}p\left(x|\theta\right)\right)^{2}}{p\left(x|\theta\right)}\,\text{dx}.
\end{equation}
Given that $p(\vec{x}|\theta)$ 
is a multivariate Gaussian distribution with mean vector $\vec{\mu}$ and covariance matrix $\Sigma$, i.e., $p(\vec{x}|\theta)\propto\exp\left[-\frac{1}{2}\left(\vec{x}-\vec{\mu}\right)^{T}\Sigma^{-1}\left(\vec{x}-\vec{\mu}\right)\right]$, then the FI about $\theta$ takes the form~\cite{kay1993statistical}: 
\begin{align}
I=\left(\partial_{\theta}\vec\mu\right)^{T}\Sigma^{-1}\left(\partial_{\theta}\vec\mu\right)+\frac{1}{2}\text{Tr}\left[\Sigma^{-1}\left(\partial_{\theta}\Sigma\right)\Sigma^{-1}\left(\partial_{\theta}\Sigma\right)\right].
\label{eq:gaussian_classical_fi}
\end{align}
The first term represents the information obtained
from the mean vector, while the second term accounts for the information contained within the covariance matrix.

In the quantum regime, the probability distribution $p(x|\theta)$ is replaced by a density matrix $\hat{\rho}(\theta)$. Consequently, the classical CRB generalizes to the quantum CRB, where the optimization extends over all possible measurements. 

The mean squared error is lower-bounded by the inverse of the QFI, $\mathcal{I}^{-1}$. 
For a density matrix $\hat{\rho}(\theta)$, the QFI is given by \cite{braunstein1994statistical}:
\begin{align}
\mathcal{I}=\sum_{j,k}\frac{2}{p_{j}+p_{k}}|\langle j|\partial_{\theta}\hat{\rho}|k\rangle|^{2},    
\end{align}
where $\{|j\rangle\}$ and $\{p_j\}$ denote the eigenstates and eigenvalues of $\hat{\rho}$, respectively (with the summation excluding terms where $p_j+p_k=0$).

For a Gaussian quantum state characterized by a displacement vector $\mu(\theta)$ and a covariance matrix $\Sigma(\theta)$, the QFI is given by~\cite{R55_Gao_2014}:
\begin{align}
\begin{split}
&\mathcal{I} = \frac{1}{2} \mathcal{M}^{-1}_{\alpha\beta, \mu\nu} \, \partial_\theta \Sigma^{\alpha\beta} \, \partial_\theta \Sigma^{\mu\nu}+\\
&\left(\partial_{\theta}\mu\right)^{T}\Sigma^{-1}\left(\partial_{\theta}\mu\right),
\label{eq:gaussian_qfi} 
\end{split}
\end{align}
where the matrix \( \mathcal{M} \) is defined as:
\begin{equation}
    \mathcal{M} = \Sigma \otimes \Sigma + \frac{1}{4} \Omega \otimes \Omega.
\label{eq:M_definition}    
\end{equation}
The classical limit (Eq.~(\ref{eq:gaussian_classical_fi})) is recovered from Eq.~(\ref{eq:gaussian_qfi}) by neglecting the $\Omega\otimes\Omega$ term in $\mathcal{M}$ \cite{monras2013phase}. 
This corresponds to the limit $\hbar\to 0$, as $\Omega$ encodes the non-commuting nature of the phase-space quadratures (see Eq.~(\ref{eq:omega})).

In the Gaussian case, it was shown in Ref.~\cite{vsafranek2015quantum} that the QFI can be decomposed into information from  the symplectic eigenvalues ,$\left\{ \nu_{i}\right\} _{i},$ and information from the symplectic transformation $M$.
To show this decomposition, let us define the generator of the symplectic transformation: $T\equiv M^{-1}\partial_{\theta}M$,
and observe that it can be written as $T=\left(\begin{array}{cc}
A & B\\
C & -A^{t}
\end{array}\right),$
with $A,B,C$ real Hermitian matrices.
Defining $\mathfrak{Q}=\frac{1}{2}\left[\left(A-A^{t}\right)+i\left(C-B\right)\right]$, $\mathfrak{Z}=\frac{1}{2}\left[\left(A+A^{t}\right)+i\left(B+C\right)\right]$,
and assuming that the parameter is encoded in the covariance matrix, the QFI can be written as follows
\begin{equation}
\begin{split}
&\mathcal{I}=\underset{i=1}{\overset{N}{\sum}}\frac{4(\partial_{\theta}{\nu}_{i})^{2}}{4\nu_{i}^{2}-1}+\\
&\underset{i,j=1}{\overset{N}{\sum}}\frac{4\left(\nu_{i}+\nu_{j}\right)^{2}}{4\nu_{i}\nu_{j}+1}|\mathfrak{Q}_{i,j}|^{2}+\underset{i,j=1}{\overset{N}{\sum}}\frac{4\left(\nu_{i}-\nu_{j}\right)^{2}}{4\nu_{i}\nu_{j}-1}|\mathfrak{Z}_{i,j}|^{2}. 
\end{split}
\label{eq:qfi_symplectic_contributions}
\end{equation}
The first term corresponds to the information from the symplectic eigenvalues and the second and third terms correspond to the information from the symplectic transformation, i.e. from the change in the normal modes.

The optimal measurement basis that saturates the QFI corresponds to the eigenbasis of the symmetric logarithmic derivative (SLD), denoted by $\mathcal{L}$. This operator is implicitly defined by the equation \cite{braunstein1994statistical}:
\begin{equation}
    \frac{1}{2}\left(\mathcal{L}\rho+\rho\mathcal{L}\right)=\partial_{\theta}\rho.
\end{equation}
For zero-mean Gaussian states (characterized by vanishing phase-space displacement), the SLD admits the following explicit representation~\cite{R55_Gao_2014}:
\begin{equation}
    \mathcal{L} = \frac{1}{2}\mathcal{M}^{-1}_{\alpha\beta,\mu\nu} \, \partial_\theta \Sigma^{\mu\nu} \left( q^\alpha q^\beta - \Sigma^{\alpha\beta} \right),
\label{eq:SLD_gaussian}    
\end{equation}
where $q^\alpha$ denotes the quadrature operators and $\mathcal{M}$ is defined in Eq.~(\ref{eq:M_definition}). Consequently, for zero-mean Gaussian states, the optimal measurement observable is a quadratic form of the quadratures, expressible as $\sum_{\mu,\nu}c_{\mu\nu}q_{\mu}q_{\nu}$.

\subsection{Fisher information for reflectivity estimation}
Using the Fisher information as the metric, we compare the ultimate quantum limit given by the QFI (denoted $\mathcal{I}$), with the precision attainable via 
optimal {\it{local}} homodyne detection. We refer to the latter as the local homodyne FI, denoted $I_\text{h,local}$.

The QFI is calculated using Eq.~(\ref{eq:gaussian_qfi})
with the TMSV covariance matrix in Eq.~(\ref{eq:Cov}),
which yields the following expression for $\mathcal{I}$
\begin{equation}
  \begin{aligned}
        &\mathcal{I} = \frac{1}{\eta(1-\eta)[(1-\eta)\tilde{N}+1]} \Bigg[n_r^2(1-\eta)
        \\&+(n_r+2\eta n_r n_\beta +\eta n_\beta)\Bigg],
    \end{aligned} 
    \label{eq:QFI}
\end{equation}
where $\tilde{N} \equiv n_\beta+n_r+2n_\beta n_r$.
We remark that 
$\mathcal{I},$ was previously obtained in Ref \cite{R36_Jonsson_2022} (up to a different parameterization; Ref \cite{R36_Jonsson_2022} considered estimation of $\sqrt{\eta}$ instead of $\eta$).

In the local homodyne strategy, we measure local quadratures of the signal and idler modes. The FI with this measurement is calculated by applying Eq.~(\ref{eq:gaussian_classical_fi}) to the projected covariance matrix (Eq.~(\ref{eq:Cov})).
Optimizing this strategy over the choice of local quadratures it can be shown that the optimal strategy is measuring the same quadratures in the signal and idler modes, e.g. measuring the $Q_i$ and $Q_s$ quadratures (see Appendix \ref{app:optimal_local_homodyne} for a short proof). The optimal local homodyne FI is then given by
\begin{align}
\begin{split}
&I_\text{h,local} = \frac{1}{\eta[2(1-\eta)\tilde{N}+1]^2}
\Bigg[
        \\&n_r(2n_\beta+1)\Big(2n_r^2(1-\eta)+3n_r+1\Big)
        \\
& +2\eta n_\beta\Big(n_r^2+n_r\Big)(4n_\beta +1)+2\eta n_\beta^2
        \Bigg].
\end{split} 
\label{eq:HFI}
\end{align}

\begin{figure*}[t]
    \centering
    \begin{tikzpicture}
        \node[anchor=south west, inner sep=0] (image) at (0,0) {
            \includegraphics[width=5.9in, trim={0mm 0mm 0mm 0mm}, clip]{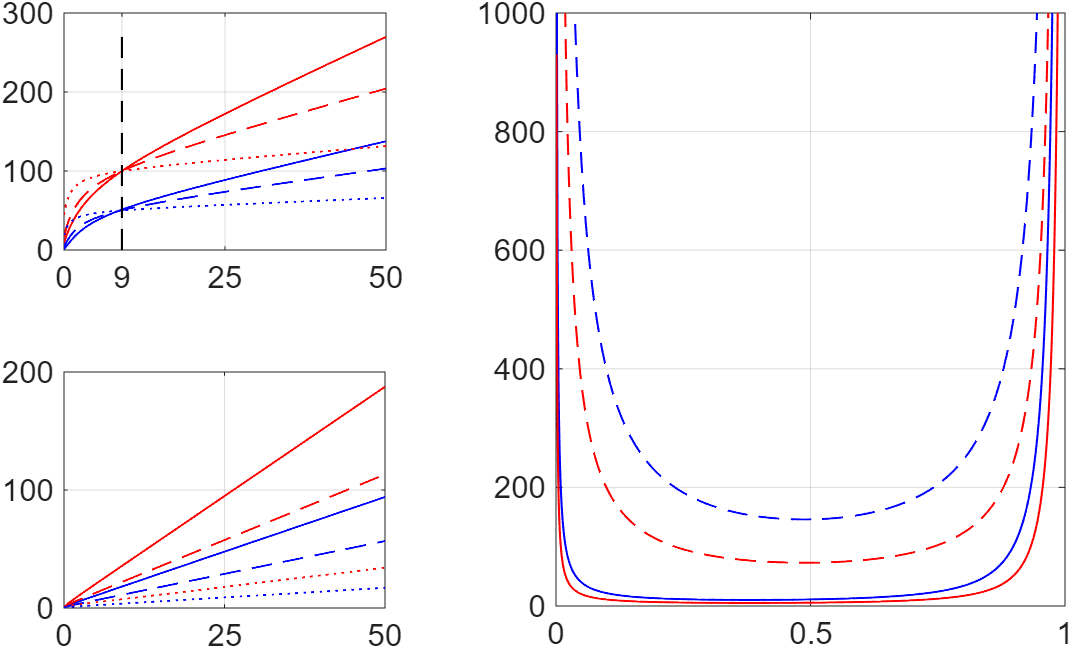}
        };

        \begin{scope}[x={(image.south east)}, y={(image.north west)}]

            \node[below,font=\Large,rotate=0] at (.45, .55) {$\mathcal{I}$};
            \node[below,font=\Large,rotate=0] at (-.02, .83) {$\mathcal{I}$};
            \node[below,font=\Large,rotate=0] at (-.02, .27) {$\mathcal{I}$};
            
            \node[below,font=\Large] at (0.76, 0) {$\eta$};
            \node[below,font=\Large] at (0.22, 0) {$n_r$};
            \node[below,font=\Large] at (0.22, .55) {$n_r$};
            \node[below,font=\normalsize] at (0.16, .94) {$n_r=n_\eta$};
            
            \node[below,font=\normalsize] at (0.08, .98) {(a)};
            \node[below,font=\normalsize] at (0.08, .43) {(b)};
            \node[below,font=\normalsize] at (0.56, .98) {(c)};

            \draw[fill=white] (.73,.836) rectangle (.85,.91);
                \node[below,font=\normalsize] at (0.79, .9) {$n_\beta\approx2.32$};

            \draw[fill=white] (.25,.896) rectangle (.33,.97);
                \node[below,font=\normalsize] at (0.29, .96) {$\eta=.9$};

            \draw[fill=white] (.25,.341) rectangle (.33,.415);
                \node[below,font=\normalsize] at (0.29, .405) {$\eta=.1$};                

            \draw[fill=white] (.68,.58) rectangle (.915,.825);
                \node[right,font=\normalsize] at (0.73, .79) {$I_\text{h,local},\ n_r=200$};
                \draw[dashed,red,thick] (0.69, .79) -- (0.73, .79);
            
                \node[right,font=\normalsize] at (0.73, .73) {$I_\text{h,local},\ n_r=10$};
                \draw[red,thick] (0.69, .73) -- (0.73, .73);
            
                \node[right,font=\normalsize] at (0.73, .67) {$\mathcal{I},\ \ n_r=200$};
                \draw[dashed,blue,thick] (0.69, .67) -- (0.73, .67);
            
                \node[right,font=\normalsize] at (0.73, .61) {$\mathcal{I},\ \ n_r=10$};
                \draw[blue,thick] (0.69, .61) -- (0.73, .61);

            \draw[fill=white] (-.23,.296) rectangle (-.01,.664);
                \node[right,font=\normalsize] at (-.18, .63) {$I_\text{h,local},\ n_\beta=8$};
                \draw[dotted,blue,thick] (-.22, .63) -- (-.18, .63);
            
                \node[right,font=\normalsize] at (-.18, .57) {$\mathcal{I},\ \  n_\beta=8$};
                \draw[dotted,red,thick] (-.22, .57) -- (-.18, .57);
            
                \node[right,font=\normalsize] at (-.18, .51) {$I_\text{h,local},\ n_\beta=2$};
                \draw[dashed,blue,thick] (-.22, .51) -- (-.18, .51);
            
                \node[right,font=\normalsize] at (-.18, .45) {$\mathcal{I},\ \ n_\beta=2$};
                \draw[dashed,red,thick] (-.22, .45) -- (-.18, .45);
            
                \node[right,font=\normalsize] at (-.185, .39) {$I_\text{h,local},\ n_\beta=1$};
                \draw[-,blue,thick] (-.22, .39) -- (-.18, .39);
            
                \node[right,font=\normalsize] at (-.18, .33) {$\mathcal{I},\ \  n_\beta=1$};
                \draw[-,red,thick] (-.22, .33) -- (-.18, .33);

        \end{scope}
    \end{tikzpicture}
    \caption{
    \label{fig:fi_plots}
      (a),(b) QFI and local homodyne FI as a function of $n_{r},$ for different values of $n_\beta$, with (a) $\eta=0.9$ and (b) $\eta=0.1.$ The dashed, black vertical line represents the crossover point of $n_r=n_\eta,$ in which the dependence on $n_\beta$ changes.
     (c) QFI and local homodyne FI as a function of $\eta,$ for $n_\beta \approx 2.32,$ and different values of $n_r$.
    }
\end{figure*}
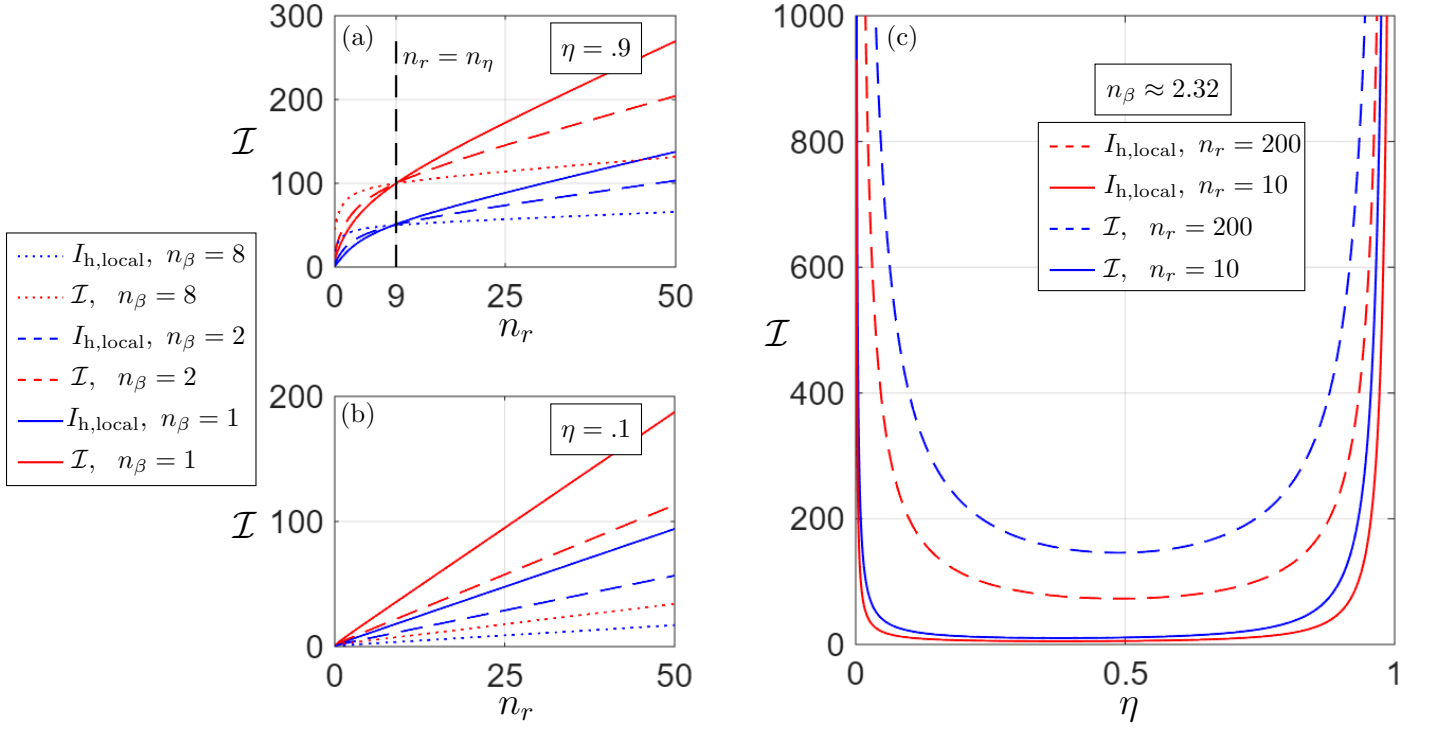
We expect $I_{\text{h,local}}$ to be non-optimal: It is later shown that local homodyne measurements do not obtain an advantage over optimal single mode strategies, and thus cannot gain from the signal-idler entanglement. This aligns with the result that local Gaussian measurements cannot achieve entanglement advantage in quantum illumination \cite{bradshaw2017overarching}.

The asymptotical behavior of $\mathcal{I}$, $I_\text{h,local}$
is summarized in table~\ref{tab:I}.

\begin{table}[H]
\centering
\resizebox{\columnwidth}{!}{
    \renewcommand{\arraystretch}{2}
    \setlength{\tabcolsep}{6pt}
    \begin{tabular}{@{}ccc@{}}
    \toprule
    $limit$ & $\mathcal{I}$ & $I_\text{h,local}$ \\
    \midrule
    $\eta\to0$   & Eq.~\ref{eq:QFI_low_reflectivity} ($\rightarrow \infty$) &    $\frac{n_{r}\left(2n_{\beta}+1\right)\left(2n_{r}^{2}+3n_{r}+1\right)}{\eta\left(2\tilde{N}+1\right)}\rightarrow\infty$      \\
    $\eta\to1$   &  Eq.~\ref{eq:QFI_high_reflectivity} ($\rightarrow \infty$)
    &     $s(n_r,n_\beta)+(n_r^2+n_r)$     \\
    \midrule
    $n_r\to0$    &     $\frac{n_\beta}{(1-\eta)[n_\beta(1-\eta)+1]}$        &    $\frac{2n_\beta^2}{[2n_\beta(1-\eta)+1]^2}$       \\
    $n_r\to \infty$    &     $\frac{n_{r}}{\eta\left(1-\eta\right)\left(1+2n_{\beta}\right)} \rightarrow \infty$        &     $\frac{n_{r}}{2\eta\left(1-\eta\right)\left(1+2n_{\beta}\right)}\rightarrow\infty$      \\
    \midrule
    $n_\beta\to0$    &      $\frac{n_r}{\eta}\cdot\frac{1}{(1-\eta)}$       &      $\frac{n_r}{\eta}\cdot\frac{[2n_r^2(1-\eta)+3n_r+1]}{[2n_r(1-\eta)+1]^2}$     \\
    $n_\beta\to \infty$    &      $\frac{1}{(1-\eta)^2}$       &     $\frac{1}{2(1-\eta)^2}$      \\
    \bottomrule
    \end{tabular}
} 
\caption{
\label{tab:I}
Asymptotic limits of $I_\text{h,local}$ and $\mathcal{I}$. Here, $s(n_r,n_\beta)=2(n_\beta+n_r)^2+8n_\beta n_r(n_\beta n_r+n_\beta + n_r)$ is a symmetric function of $n_r$ and $n_\beta$.
}
\end{table}

\subsubsection{Dependence on $n_r$}
As expected, both $\mathcal{I}$ and $I_\text{h,local}$ grow monotonically with $n_r$: increasing the photon number (or squeezing) yields more information. This behavior is illustrated in Fig.~\ref{fig:fi_plots}(a-b).
Specifically, for sufficiently large $n_r$, we get
\begin{align}
\mathcal{I}\approx\frac{1}{\eta (1-\eta)(1+2n_\beta)}n_r,
\label{eq:qfi_large_nr}
\end{align}
and $I_\text{h,local}=\frac{1}{2}\mathcal{I}$. Consequently, both quantities diverge as $n_r \rightarrow \infty$.

Fig.~\ref{fig:fi_plots}(b) shows that for low reflectivities, the information follows this linear trend even at small $n_r$. However, for high-reflectivity targets (Fig.~\ref{fig:fi_plots}(a)), the behavior is more complex. In the high $\eta$ regime, $\mathcal{I}$ can be approximated as:
\begin{align}
\mathcal{I} \approx \frac{n_{r}^{2}}{\left(1-\eta\right)\tilde{N}}+\frac{\tilde{N}}{\left(1-\eta\right)\left[(1-\eta)\tilde{N}+1\right]}
\label{eq:QFI_high_reflectivity}
\end{align}
where $\tilde{N}=n_{\beta}+n_{r}+2n_{\beta}n_{r}$.
At small $n_r$, the second term in Eq.~(\ref{eq:QFI_high_reflectivity}) dominates. This term does not grow linearly but converges to $1/(1-\eta)^2$. For larger $n_r$, the first term becomes dominant, restoring the asymptotic linear behavior. In Fig.~\ref{fig:fi_plots}(a), the transition between these two regimes is illustrated by the dashed black line. This transition occurs at $n_r=n_\eta$ and will be further discusses in section \ref{I_nb}. It will be later shown that this point also marks the boundary between optimal measurement strategies.

We now turn to the $n_r \to 0$ limit, representing the single-mode vacuum case. The relevant $\mathcal{I},$ $I_\text{h,local}$ expressions are given in Table~\ref{tab:I} (3rd row). It can be seen that the low and high reflectivity regimes are very different. In the low reflectivity regime, $\eta \rightarrow 0,$ both $\mathcal{I}$ and $I_\text{h,local}$ are finite, and the ratio between them is limited: $\frac{I_\text{h,local}}{\mathcal{I}}=\frac{1}{2}\left[1- \frac{1}{(2n_\beta(1-\eta)+1)^2} \right]\leq\frac{1}{2}$.

On the other hand, in the high reflectivity regime, $\eta \rightarrow 1$, the ratio between $\mathcal{I}$ and $I_\text{h,local}$ can be very large: as $\eta \rightarrow 1$, $\mathcal{I}$ diverges as $\frac{n_{\beta}}{1-\eta}$, while the local homodyne FI remains finite: $I_{h,local}=2 n_\beta^2$. This stark difference was observed in equivalent sensing scenarios, particularly in stochastic signal sensing~\cite{ng2016spectrum, tsang2023quantum, R65_shi2023ultimate, gardner2025stochastic, gardner2025lindblad}.
The intuition behind this large gap is that the limit $\eta \rightarrow 1$ is equivalent to sensing a small added noise, much smaller than the vacuum noise. In homodyne measurement, this small noise is overshadowed by the vacuum noise, whereas with number-resolving measurement, we can overcome this limitation.

\subsubsection{Dependence on $\eta$}
The dependence on reflectivity is illustrated in Fig.~\ref{fig:fi_plots}(c), where we plot both $\mathcal{I}$ and $I_\text{h,local}$ as functions of $\eta$. It is evident that $\mathcal{I} \rightarrow \infty$ in both the low (Eq.~(\ref{eq:QFI_low_reflectivity})) and high (Eq.~(\ref{eq:QFI_high_reflectivity})) reflectivity limits.

However, the behavior of $\mathcal{I}$ differs significantly between these two regimes. In the high reflectivity limit, $\mathcal{I}$ diverges for every choice of $n_\beta$ and $n_r$. Conversely, in the low reflectivity limit, $\mathcal{I}$ diverges only when $n_r > 0$; for $n_r = 0$, $\mathcal{I}$ remains finite. This can be seen from the approximation of $\mathcal{I}$ in the low reflectivity limit:
\begin{align}
\mathcal{I} \approx \frac{n_{\beta}}{n_{\beta}+\frac{1+n_{r}}{1+2n_{r}}} + \frac{1}{\eta}\frac{n_{r}}{1+n_{\beta}\frac{1+2n_{r}}{1+n_{r}}}.
\label{eq:QFI_low_reflectivity}
\end{align}
$I_\text{h,local}$ exhibits behavior similar to $\mathcal{I}$ in the low reflectivity regime, where it also diverges only for $n_r > 0$ (1st row of Table~\ref{tab:I}). However, unlike $\mathcal{I}$, in the high reflectivity regime $I_\text{h,local}$ converges to a constant value (assuming fixed $n_r$ and $n_\beta$), as shown in Table~\ref{tab:I} (2nd row).

\subsubsection{Dependence on $n_\beta$}
\label{I_nb}
The effect of $n_\beta$ depends on the regime set by $n_r$ and $\eta.$ The point $n_r=n_\eta$ marks a crossover:
For $n_r>n_\eta,$ increasing $n_\beta$ 
lowers the QFI, while for $n_r <n_\eta,$ the effect is the opposite: the QFI is increased with $n_\beta$. At 
the crossover, the QFI is independent of $n_\beta$ and equals to $\mathcal{I}=1/\left( 1-\eta \right)^2.$ A similar crossover appears also for $I_\text{h,local}$ at a different point. This behavior is illustrated in Fig.~\ref{fig:fi_plots}(a).  

Let us consider also the extreme limits of $n_\beta \to 0, \infty$.
For $n_\beta \to 0$ the QFI reduces to
$\mathcal{I}=\frac{n_r}{\eta \left(1-\eta \right)}.$
The local homodyne FI, $I_\text{h,local},$ approaches this bound at low $\eta,$ but remains finite at $\eta \to 1$, failing to achieve $1/\left( 1-\eta \right)$ divergence.
The large thermal noise limit $n_\beta \to \infty$ (6th row of Table~\ref{tab:I})
requires more care since the order of limits matters: the limits of $n_\beta \to \infty $ and $\eta \to 0$ do not commute. 
Taking first the limit of $n_\beta \to \infty,$ both $\mathcal{I}$ and $I_\text{h,local}$ scale as $1/(1-\eta)^2$,
yielding little information at low reflectivities.

However, this limit assumes $\eta n_\beta \gg n_r^2 \left( 1-\eta\right), n_r,$ so $\eta$ must approach zero slowly enough.
If instead $\eta\lesssim\text{max}\{n_r, n_r^2 \}/n_\beta,$ then the usual $1/\eta$ divergence of the FIs is retrieved.
The intuition is that if $\eta n_\beta$ is large enough 
the number of signal photons is negligible compared to the noise photons.

\subsection{Comparison with coherent state probe}

We now compare the TMSV information quantities $\mathcal{I}$ and $I_\text{h,local}$ to the information obtained using a single mode coherent state probe  $|\alpha\rangle$ with the same average number of photons $|\alpha|^2=n_r$.

Without loss of generality, we can assume a real $\alpha,$ such that the initial displacement vector is $\vec\mu=(\sqrt{2}\alpha,0)^{t}.$ 
Following the encoding channel, the displacement vector and covariance matrix become $\vec\mu_{\eta}=(\sqrt{2\eta}\alpha,0)^{t}$ and $\Sigma_{\eta}=\left[1/2+(1-\eta)n_{\beta}\right]\mathbb{I}$, respectively.
Consequently, the QFI is given by:
\begin{align}
\mathcal{I}_\text{coh} =
\frac{n_{\beta}}{\left(1-\eta\right)\left[n_{\beta}\left(1-\eta\right)+1\right]}
+
\frac{n_{r}}{\eta\left[2n_{\beta}\left(1-\eta\right)+1\right]}
\label{eq:QFI_coherent}
\end{align}
The second term scales linearly with $n_r$ and corresponds to the information obtained from the displacement vector. The first term represents the vacuum QFI (the limit as $n_r\to 0$) and corresponds to the information obtained from the covariance matrix.
The homodyne FI for the coherent state is given by:
\begin{align}
I_\text{coh} = \frac{2n_{\beta}^{2}}{\left[2n_{\beta}\left(1-\eta\right)+1\right]^{2}} + \frac{n_{r}}{\eta\left[2n_\beta\left(1-\eta\right)+1\right]}.
\label{eq:FI_homodyne_coherent}
\end{align}
The second term in  Eq.~\eqref{eq:FI_homodyne_coherent} is identical to the displacement term in $\mathcal{I}_\text{coh}$. Thus, in the limit of large photon number, the homodyne FI saturates the Quantum Cramer-Rao bound:
\begin{align}
\lim_{n_{r}\rightarrow\infty} I_\text{coh} \rightarrow \mathcal{I}_\text{coh}
\end{align}
However, the first term (vacuum contribution) in $I_\text{coh}$ is strictly smaller than the corresponding vacuum term in $\mathcal{I}_\text{coh}$.

\begin{figure*}[t]
    \centering
    \begin{tikzpicture}

        \node[anchor=south west, inner sep=0] (image) at (0,0) {
            \includegraphics[width=6.5in, trim={0mm 0mm 0mm 0mm}, clip]{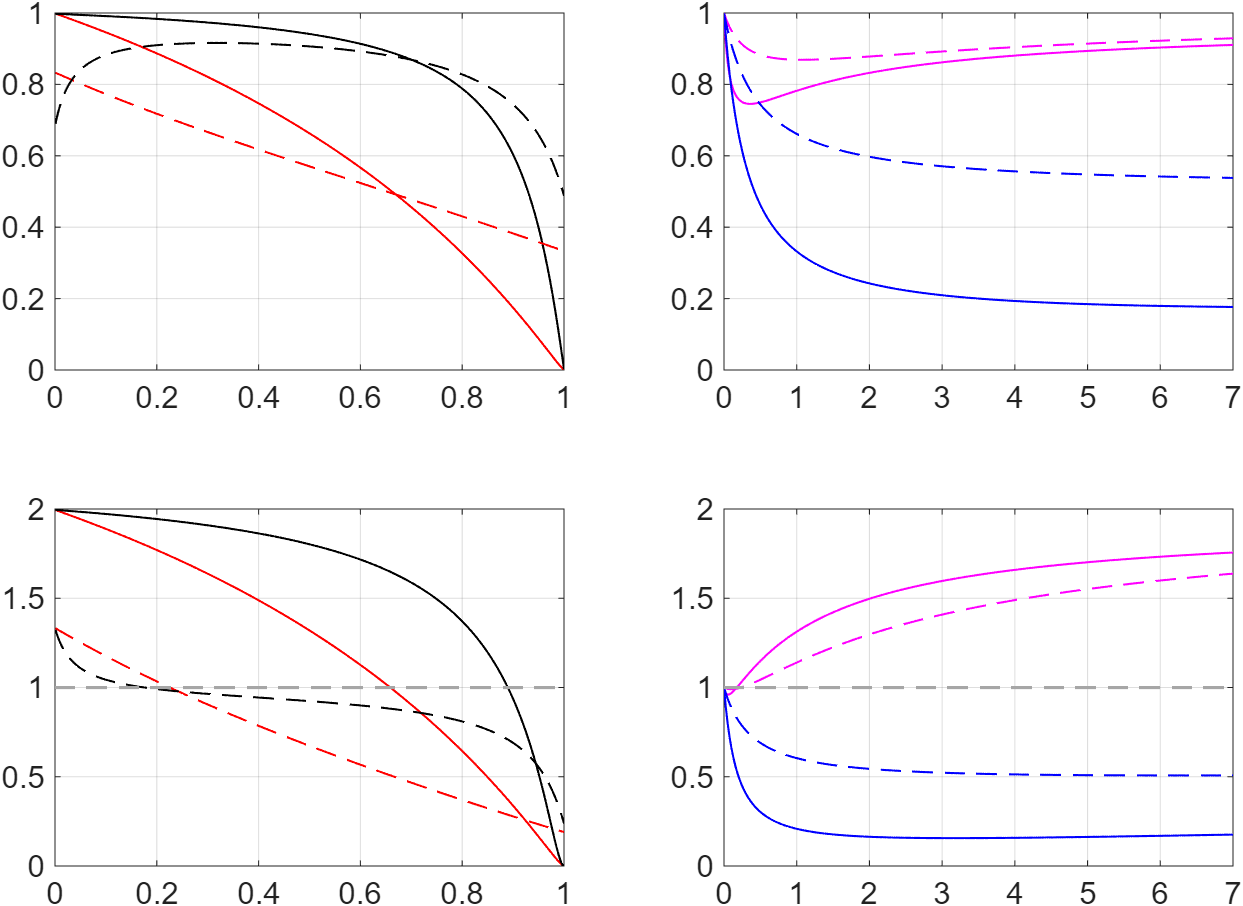}
        };

        \begin{scope}[x={(image.south east)}, y={(image.north west)}]

            \node[below,font=\Large] at (-.03, .82) {$\mathcal{R}$};
            \node[below,font=\Large] at (0.25, .54) {$\eta$};

            \node[below,font=\Large] at (.51, .82) {$\mathcal{R}$};
            \node[below,font=\Large] at (0.78, .54) {$n_r$};
            
            \node[below,font=\Large] at (-.03, .28) {$R$};
            \node[below,font=\Large] at (0.25, 0.01) {$\eta$};

            \node[below,font=\Large] at (.51, .28) {$R$};
            \node[below,font=\Large] at (0.78, 0.01) {$n_r$};

            \node[left,font=\normalsize] at (.082, .61) {(a)};
            \node[left,font=\normalsize] at (.62, .61) {(b)};
            \node[left,font=\normalsize] at (.082, .062) {(c)};
            \node[left,font=\normalsize] at (.62, .42) {(d)};

            \draw[fill=white] (0.032,-.1) rectangle (.465,-.245);
            
                \node[right,font=\normalsize] at (.07,-.14) {$n_r=200, n_\beta=10$};
                \draw[-,black,thick] (.04,-.14) -- (.07,-.14);

                \node[right,font=\normalsize] at (.07,-.2) {$n_r=.5, n_\beta=10$};
                \draw[dashed,black,thick] (.04,-.2) -- (.07,-.2);

                \node[right,font=\normalsize] at (.29,-.14) {$n_r=200, n_\beta=.5$};
                \draw[-,red,thick] (.26,-.14) -- (.29,-.14);

                \node[right,font=\normalsize] at (.29,-.2) {$n_r=.5, n_\beta=.5$};
                \draw[dashed,red,thick] (.26,-.2) -- (.29,-.2);
                
            \draw[fill=white] (0.56,-.1) rectangle (.993,-.245);
            
                \node[right,font=\normalsize] at (.598,-.14) {$\eta=.9, n_\beta=1$};
                \draw[-,blue,thick] (.568,-.14) -- (.598,-.14);

                \node[right,font=\normalsize] at (.598,-.2) {$\eta=.9, n_\beta=10$};
                \draw[dashed,blue,thick] (.568,-.2) -- (.598,-.2);

                \node[right,font=\normalsize] at (.828,-.14) {$\eta=.1, n_\beta=1$};
                \draw[-,draw={rgb:RubineRed,.9;VioletRed,1},thick] (.828,-.14) -- (.798,-.14);

                \node[right,font=\normalsize] at (.828,-.2) {$\eta=.1, n_\beta=10$};
                \draw[dashed,draw={rgb:RubineRed,.9;VioletRed,1},thick] (.828,-.2) -- (.798,-.2);

        \end{scope}
    \end{tikzpicture}
    \caption{
    \label{fig:R}
    (a) The ratio $\mathcal{R} = \mathcal{I}_\text{coh}/\mathcal{I}$ as a function of reflectivity $\eta$, (b) $\mathcal{R}$ as a function of $n_r$.
    The TMSV QFI is always larger, and thus $\mathcal{R} \leq 1,$ where the maximal gap is at larger $\eta.$
    (c)  $R = I_\text{coh}/I_\text{h,local}$ as a function of reflectivity $\eta$ (d) and mean  photon number $n_r$. The gray line corresponds to $R=1.$
    }
\end{figure*}
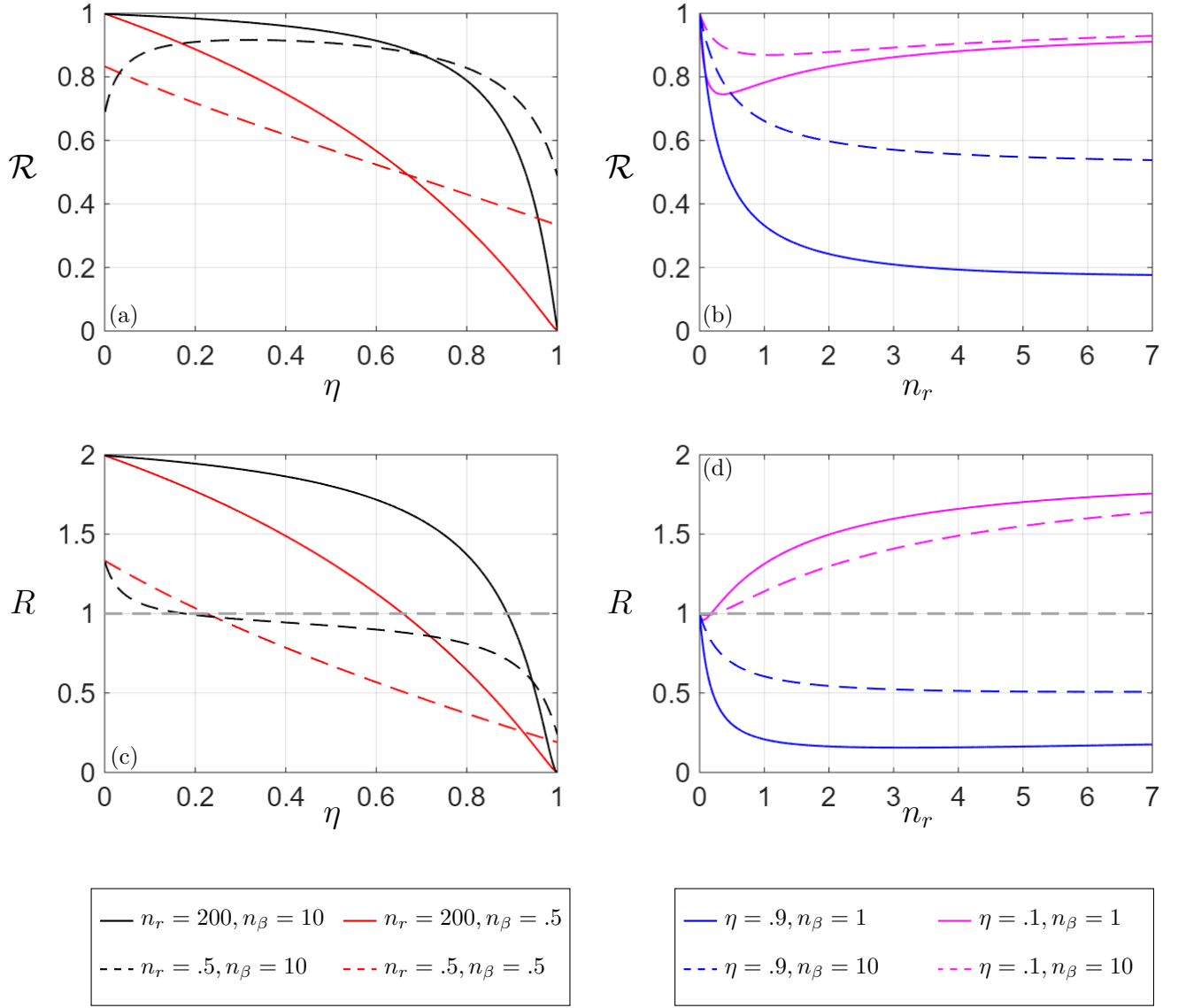

We denote the ratios between the coherent state (Q)FI and the TMSV (Q)FI as
$\mathcal{R} = \mathcal{I}_\text{coh}/\mathcal{I}$ and $R = I_\text{coh}/I_\text{h,local}.$
The asymptotic behavior of theses ratios
in various limits is summarized in Table~\ref{tab:II}.
\begin{table}[H]
\centering
\renewcommand{\arraystretch}{2}
\setlength{\tabcolsep}{6pt}
\begin{tabular}{@{}ccc@{}}
\toprule
$limit$ & $\mathcal{R}$ & $R$ \\
\midrule
$\eta\to0$ &
$1-\frac{1}{n_r+1}\frac{n_\beta}{2n_\beta+1}$ &
$1+\frac{1}{n_r+1}$
\\
$\eta\to1$ &
$\frac{n_\beta}{\tilde{N}}$ &
$\frac{2n_\beta^2+n_r}{2 \tilde{N}^2+(n_r^2+n_r)}$
\\
\midrule
$n_r\to0$ &
$1$ &
$1$
\\
$n_r\to \infty$ &
$1-\frac{\eta}{2n_\beta(1-\eta)+1}$ &
$2\left[1-\frac{\eta}{2n_\beta(1-\eta)+1}\right]$
\\
\midrule
$n_\beta\to0$ &
$1-\eta$ &
$2(1-\eta)\left[1+\frac{\frac{1}{2(1-\eta)}-(n_r+1)}{2n_r^2(1-\eta)+3n_r+1}\right]$
\\
$n_\beta\to \infty$ &
$1$ &
$1$
\\
\bottomrule
\end{tabular}
\caption{
\label{tab:II}
Asymptotic limits of the ratios $\mathcal{R}=\mathcal{I}_\text{coh}/\mathcal{I}$ and $R=I_\text{coh}/I_\text{h,local}$, where $\tilde{N} = n_r+n_\beta + 2n_r n_\beta$.
}
\end{table}
It is well established that the TMSV QFI ($\mathcal{I}$) outperforms the coherent state QFI ($\mathcal{I}_\text{coh}$) for any given $n_r,n_\beta$ and $\eta$, i.e. $\mathcal{R}\leq1$ \cite{R36_Jonsson_2022} (for completeness we also show this in appendix~\ref{apx:Rgeq1}).
This implies that assuming we can perform the optimal general measurement, the TMSV probe consistently yields a metrological advantage over the coherent probe. This advantage is further illustrated in Fig.~\ref{fig:R}(a-b) and in Table~\ref{tab:II}.

In Fig.~\ref{fig:R}(a), we analyze the dependence on reflectivity. In the high squeezing scenario (solid lines), $\mathcal{R}$ approaches 1 for low reflectivities, indicating that the coherent probe yield approximately the same amount of information as the TMSV probe. As $\eta$ increases, $\mathcal{R}$ decreases indicating the TMSV probe outperform the coherent probe.

For large $n_\beta$ (black solid line), $\mathcal{R}$ remains relatively flat (above $\approx 0.9$) for reflectivities up to $\eta \approx 0.7$; for higher reflectivities, the slope increases strictly, indicating a rapid gain in quantum advantage. Specifically, as $\eta\to1$, $\mathcal{R}$ approaches zero (4th,5th rows of Table~\ref{tab:II}). This indicates that when the signal energy significantly dominates the thermal noise, the TMSV probe yields a diverging advantage over the coherent probe.
A special case arises at $n_\beta=0$ (5th row of Table~\ref{tab:II}); here, the choice of $n_r$ is irrelevant for the ratio.

In the low squeezing scenario (dashed lines), we observe that for low reflectivities, $\mathcal{R}$ decreases as the noise increases. This behavior can also be seen in the 1st row of Table~\ref{tab:II}. This suggests that in the low-reflectivity limit, the TMSV probe performance is more robust against noise compared to the coherent probe performance. Conversely, at high reflectivities, the opposite trend holds: $\mathcal{R}$ increases with $n_\beta$ (2nd row of Table~\ref{tab:II}), implying that the coherent probe is less affected by noise than the TMSV probe.

Fig.~\ref{fig:R}(b) displays the dependence on $n_r$. As $n_r \to \infty$, the ratio $\mathcal{R}$ is given by $\mathcal{R} = 1-\frac{\eta}{2n_\beta(1-\eta)+1}$. As a result, the dashed curves (high noise) are consistently positioned above the solid lines (low noise). Therefore, higher thermal noise thereby increasing the relative effectiveness of the coherent probe compared to the TMSV probe. In other words, the quantum advantage of the TMSV state is more pronounced in low-noise environments.

Furthermore, in the high-squeezing regime, the deviation between the high-noise (dashed) and low-noise (solid) curves diminishes for low reflectivities. We can see that the low reflectivity magenta curves are closer to each other than the high reflectivity blue curves. This suggests that in the low-$\eta$ limit, the performance ratio $\mathcal{R}$ is largely insensitive to thermal noise. Whereas for higher reflectivities, a pronounced separation emerges, indicating a strong dependence on the noise level $n_\beta$.

Under the restriction of local homodyne measurement, the landscape changes significantly. 

Here, the TMSV probe performs worse than classical coherent state at low $\eta$, but regains an advantage at larger $\eta$. This can be seen from examining the two extreme $\eta$ limits in Table~\ref{tab:II}: as $\eta \to 0$ (1st row), the ratio $R$ satisfies $1\leq R\leq 2$. This implies that for low reflectivities, the coherent probe can extract up to twice the amount of information as the TMSV probe.
Conversely, as $\eta \to 1$ (2nd row), we find $R \leq 1$, indicating that the TMSV probe regains the advantage. There is therefore a crossover reflectivity $\eta_{eq}$ such that $R(\eta_{eq}) = 1$, which is illustrated in Fig.~\ref{fig:R}(c). This crossover point depends on $n_r,n_\beta,$ and in the large $n_r$ limit it is given by
\begin{equation}
    \eta > \eta_{eq} \equiv \frac{2n_\beta+1}{2n_\beta+2} \quad \Leftrightarrow \quad n_\eta > n_{\eta_{eq}} \equiv 2n_\beta+1.
    \label{eq:adv}
\end{equation}
Recall the entanglement criterion from Eq.~(\ref{eq:Dn}): $n_\eta > n_\beta$. Comparing this with Eq.~\eqref{eq:adv}, we conclude that for the TMSV probe to provide an advantage under local homodyne detection in the high $n_r$ regime, the 
state after the encoding must be entangled. Moreover, observing Eq.~\eqref{eq:adv}, we see that $\partial_{n_\beta}\eta_{eq} = 0.5(n_\beta+1)^{-2}\geq0$. This implies that for large $n_r$, the threshold $\eta_{eq}$ increases with $n_\beta$, i.e the region of coherent probe advantage expands, as shown by the solid lines in Fig.~\ref{fig:R}(c). However, for small $n_r$ (dashed lines), the behavior is reversed: $\eta_{eq}$ decreases as $n_\beta$ increases.

The dependence of $R$ on $n_r$ is illustrated in Fig.~\ref{fig:R}(d).
For small $\eta$ (magenta lines), the coherent state probe performs better for almost all values of $n_r$ (and larger $n_r$ only increases the gap), whereas for larger $\eta$ (blue lines), the TMSV probe is better across the entire $n_r$ range.

The analysis above shows that given the constraint of local homodyne measurements, a TMSV probe provides no advantage in the low-reflectivity regime. The reason for this inefficiency is that TMSV probe with local homodyne measurements is effectively equivalent to a single-mode strategy. In Appendix~\ref{apx:local_homodyne_no_entanglement}, we show that this measurement scheme is equivalent to performing homodyne on a single mode squeezed state with a mean vector of $\vec{\mu}=\left(\begin{array}{cc}\sqrt{\frac{2n_{r}\left(n_{r}+1\right)}{2n_{r}+1}} & 0\end{array}\right)^{t},$ a covariance matrix of $V=\text{diag}\left\{ \frac{1}{4n_{r}+2},4n_{r}+2\right\},$ and mean photon number $n_r.$ Hence this local homodyne strategy cannot obtain an entanglement advantage. Furthermore, in the low-$\eta$ regime, the optimal single-mode probe is known to be a coherent state~\cite{R25_PhysRevLett.118.070803}, which outperforms any single-mode squeezed state in this limit. 
For larger values of $\eta,$ the coherent state is no longer the optimal single-mode state, and the corresponding squeezed-state strategy can indeed outperform it; however, it still does not saturate the TMSV QFI. This motivates the search for other practical Gaussian measurements that perform better 
and can be nearly-optimal.
In the next section we find general optimal measurements and show that nearly optimal precision can be obtained using {\it{non-local}} homodyne measurements.

\section{Measurement Strategy\label{sec:IV}}
In this section, we derive optimal measurements that saturate the QFI and study the performance of optimized homodyne detection schemes.
The performance of optimal Gaussian detection schemes is particularly relevant for 
the microwave domain in which 
practical readout schemes are
typically restricted to Gaussian measurements \cite{Nori}.

\subsection{
Optimal Measurement Strategy\label{sec:IV.A}}
The optimal measurement is derived from the SLD via Eq.~(\ref{eq:SLD_gaussian}). In our case, the SLD takes the form:
\begin{equation}
\begin{aligned}
    \mathcal{L} &= -\frac{1}{1-\eta}\Bigg[\frac{1}{ (1-\eta) \tilde{N}+1}\left(\sum_{i,j=1}^4c_{ij} q_i q_j\right) - 1\Bigg],\\
    &\text{with coefficients:}\\
    &\ \ c_{ii} \equiv c_d = n_r+\frac{1}{2},\\
    &\ \ c_{24} = c_{42} = -c_{13} = -c_{31}\equiv c_{od} = \frac{1+\eta}{2\sqrt{\eta}}\sqrt{n_r^2+n_r}, \\
    &\ \ c_{\text{else}} = 0.
\end{aligned}   
\label{eq:SLD}
\end{equation}
Neglecting constant factors and terms that are proportional to the identity,
the optimal observable can be expressed as:
\begin{equation}
\mathcal{L} \propto c_{d}\sum_{\mu=i,s}^{ }\left(Q_{\mu}^{2}+P_{\mu}^{2}\right) + 2c_{od}\left(P_{i}P_{s}-Q_{i}Q_{s}\right).
\label{eq:sld_pa_form}
\end{equation}
This expression corresponds to a quadratic form of the quadratures 
given
by the block matrix $\mathcal{M}_{\mathcal{L}}$:
\begin{equation}
    \mathcal{M}_\mathcal{L} = 
    \begin{bmatrix}
     c_{d}\hat{\mathbb{I}}_2  & -c_{od}\hat{\sigma}_z \\
     -c_{od}\hat{\sigma}_z  & c_{d}\hat{\mathbb{I}}_2\\ 
    \end{bmatrix},
\end{equation}
where $\hat{\mathbb{I}}_2$ is the $2\times2$ identity matrix and $\hat{\sigma}_z$ is the Pauli-Z matrix acting on the quadrature subspace. Diagonalizing this quadratic form we obtain that $\mathcal{L}$ can be written as 
\begin{align}
\mathcal{L}\propto c_{+}\left(P_{+}^{2}+Q_{-}^{2}\right)+c_{-}\left(P_{-}^{2}+Q_{+}^{2}\right),
\label{eq:sld_diagonal}    
\end{align}
where $P_{\pm}=\frac{1}{\sqrt{2}}\left(P_{i}\pm P_{s}\right),$ $Q_{\pm}=\frac{1}{\sqrt{2}}\left(Q_{i}\pm Q_{s}\right),$ and
\begin{equation}
\begin{aligned}
    c_\pm = c_d\pm c_{od} &= \left(n_r+\frac{1}{2}\right) \pm \frac{1+\eta}{2\sqrt{\eta}}\sqrt{n_r^2+n_r}    
    \\&=
    \left(n_r+\frac{1}{2}\right) \pm \left(n_\eta+\frac{1}{2}\right)\sqrt{\frac{n_r^2+n_r}{n_\eta^2+n_\eta}}.
\end{aligned}
\label{eq:c_pm}
\end{equation}
Let us also define the $r_s$ coefficient:
\begin{equation}
    r_s \equiv -\frac{1}{4}\ln\left(\frac{|c_-|}{c_+}\right)
    =
    \left\{
    \begin{matrix}
        \frac{1}{2}\arctanh\left(\frac{c_{od}}{c_{d}}\right) & c_->0 \\
        \frac{1}{2}\arctanh\left(\frac{c_d}{c_{od}}\right) & c_-<0,
    \end{matrix}
    \right.
\label{eq:r_s}
\end{equation}
which will be used later.

The sign of $c_-$ determines three different SLD regimes. First, the \textbf{reflectivity dominated regime} ($c_- > 0$) corresponds to $n_r < n_\eta$ or equivalently $\eta > \eta^*$, where $\eta^* = n_r/(n_r+1)$. Second, the \textbf{squeezing dominated regime} ($c_- < 0$) corresponds to $n_r > n_\eta$ ($\eta < \eta^*$). Finally, the \textbf{transition point} ($c_- = 0$) occurs at $n_r = n_\eta$ ($\eta = \eta^*$).

\subsubsection{Reflectivity Dominated Regime ($\eta>\eta^*$)}
In this regime, the optimal observable $\mathcal{L}$ can be realized by applying a parametric amplifier followed by photon number resolving measurements.

To show this, we observe that the SLD can be mapped to the total number operator through a parametric-amplifier transformation. In the Heisenberg picture, the action of the parametric amplifier unitary $\hat{S}(r')$ (Eq.~(\ref{eq:Pamp})) on the total energy observable $\sum_{\mu} (Q_{\mu}^{2}+P_{\mu}^{2})$ yields:
\begin{align}
\begin{split}
&\hat{S}(r')^{\dagger}\left(\sum_{\mu=i,s}Q_{\mu}^{2}+P_{\mu}^{2}\right)\hat{S}(r')=\\
&\cosh\left(2r'\right)\left(\sum_{\mu=i,s}Q_{\mu}^{2}+P_{\mu}^{2}\right)+2\sinh\left(-2r'\right)\left(P_{i}P_{s}-Q_{i}Q_{s}\right).
\end{split}
\end{align}
Comparing this to the expression of $\mathcal{L}$ in Eq.~(\ref{eq:sld_pa_form}), we observe that for $\eta>\eta^{*}$, the SLD takes an identical form up to a constant factor: $\mathcal{L} \propto \cosh(2r_s)(\sum Q_i^2 + P_i^2) + 2\sinh(2r_s)(P_i P_s - Q_i Q_s)$.
Hence the SLD observable can be measured by applying a parametric amplifier $\hat{S}(r')$
with $r'=-r_s$ followed by a number resolving measurement.

We note that in the limit of $\eta \to 1$, the optimal parametric amplifier receiver becomes the inverse of the one used to prepare the TMSV state, i.e. an echo protocol is optimal. To see this observe that for $\eta=1$, $r_s=\frac{1}{2}\arctanh\left(\frac{\sqrt{n_{r}^{2}+n_{r}}}{n_{r}+1/2}\right)=r$ (Eq.~(\ref{eq:cov_tmsv})).
Hence the required parametric amplifier unitary is $\hat{S}(-r),$ which is the inverse of $\hat{S}(r)$ used to generate the state.

We remark that in this regime the SLD can also be written as
\begin{align}
\mathcal{L}\propto
\left( Q_{-}'^2+P_{-}'^2\right)
+\left(Q_{+}'^2+ P_{+}'^2\right) 
\label{SLD:PA_rec}
\end{align}
where 
$Q'_{\pm},P'_{\pm}$ are the following squeezed quadratures: $\ensuremath{P'_{\pm}=e^{\pm2r_{s}}P_{\pm}}$,$\ensuremath{Q'_{\pm}=e^{\mp2r_{s}}Q_{\pm}}$.
Hence the SLD corresponds to a number resolving measurement of the $a'_{+}, a'_{-}$ modes. An alternative implementation is thus using balanced beam splitter followed by squeezing of the output modes.

Finally, we remark that implementing this method with a parametric amplifier requires precise knowledge of $r_s$, which depends on the unknown parameter $\eta$. Therefore, attaining the quantum limit in practice may require an adaptive measurement scheme.

\subsubsection{Squeezing Dominated Regime $(\eta < \eta^{*})$}
The SLD in this regime can be written as $\mathcal{L}\propto e^{2r_{s}}\left(P_{+}^{2}+Q_{-}^{2}\right)-e^{-2r_{s}}\left(P_{-}^{2}+Q_{+}^{2}\right)$, hence defining the squeezed quadratures $P'_{\pm}=e^{\pm2r_{s}}P_{\pm}$, $Q'_{\pm}=e^{\mp2r_{s}}Q_{\pm},$
the SLD reduces to
\begin{align}
\mathcal{L}\propto \left( Q'{}_{-}^{2}-P'{}_{-}^{2} \right)-\left(Q'{}_{+}^{2}- P'{}_{+}^{2} \right).
\label{SLD:SQ}
\end{align}
The optimal observable thus corresponds to  the squeezing generators of the $\hat{a}'_{+},\hat{a}'_{-}$ modes.
The spectrum of this SLD is continuous, and thus in principle cannot be realized by a Gaussian unitary followed by a number resolving measurement of the modes. 
However, in certain limits this 
measurement can be approximated using parametric amplifier receivers.
For example, in the quantum illumination limit of  $\eta \to 0$, $n_r \rightarrow0,$
the SLD reduces to $\mathcal{L} \propto P_iP_s-Q_iQ_s$.
In this limit, a suitable parametric amplifier receiver
can approximate this observable and nearly saturate the QFI \cite{guha2009gaussian,R17_Assouly2023}. 
We further analyze the optimal parametric amplifier FI in Appendix~\ref{app:optimal_parametric_amplifier} and show that there is typically a gap between this FI and the QFI, which depends on the values of $n_r,n_\beta.$
Hence, while this receiver can approach the QFI also in this regime, it is no longer optimal.

\subsubsection{Transition point ($\eta=\eta^{*}$)}
The transition point is given by  $c_-=0$, or equivalently $\eta=\eta^{*}$. Since $c_-=0$ we obtain from Eq.~(\ref{eq:sld_diagonal}) that the SLD reduces to:
\begin{equation}
    \mathcal{L} \propto c_+\Big(P_+^2 + Q_-^2\Big).
\end{equation}
The information is thus entirely contained in the $P_+$ and $Q_-$ quadratures. Since these operators commute ($[P_+, Q_-] = 0$), the optimal strategy consists of simultaneous measurements of these  quadratures.
This can be implemented by mixing the modes on a balanced beam splitter followed by standard local homodyne detection.

We remark that this measurement strategy corresponds to the double homodyne scheme proposed in Ref.~\cite{R37_PhysRevResearch.3.013006}. Our analysis shows that this scheme is strictly optimal at the transition point $n_r=n_\eta$. Furthermore, for $n_\eta > n_r \geq 0.1$ we obtain that $c_+ \gg |c_{-}|$. In this domain, $\mathcal{L}$ remains dominated by the term $P_+^2 + Q_-^2$, which implies that this double homodyne measurement should be nearly optimal in this regime. This behavior is analyzed further in Sec.~\ref{susec:homodyne_meas_strategy}.
However, implementing this scheme experimentally introduces physical challenges. Any joint operation on the signal and idler modes requires compensating for the propagation delay. This synchronization can be challenging if the range to the target
is unknown.

\begin{figure}[H]
    \centering
    \includegraphics[scale=0.56,trim={0mm 0mm 0mm 0mm}, clip]{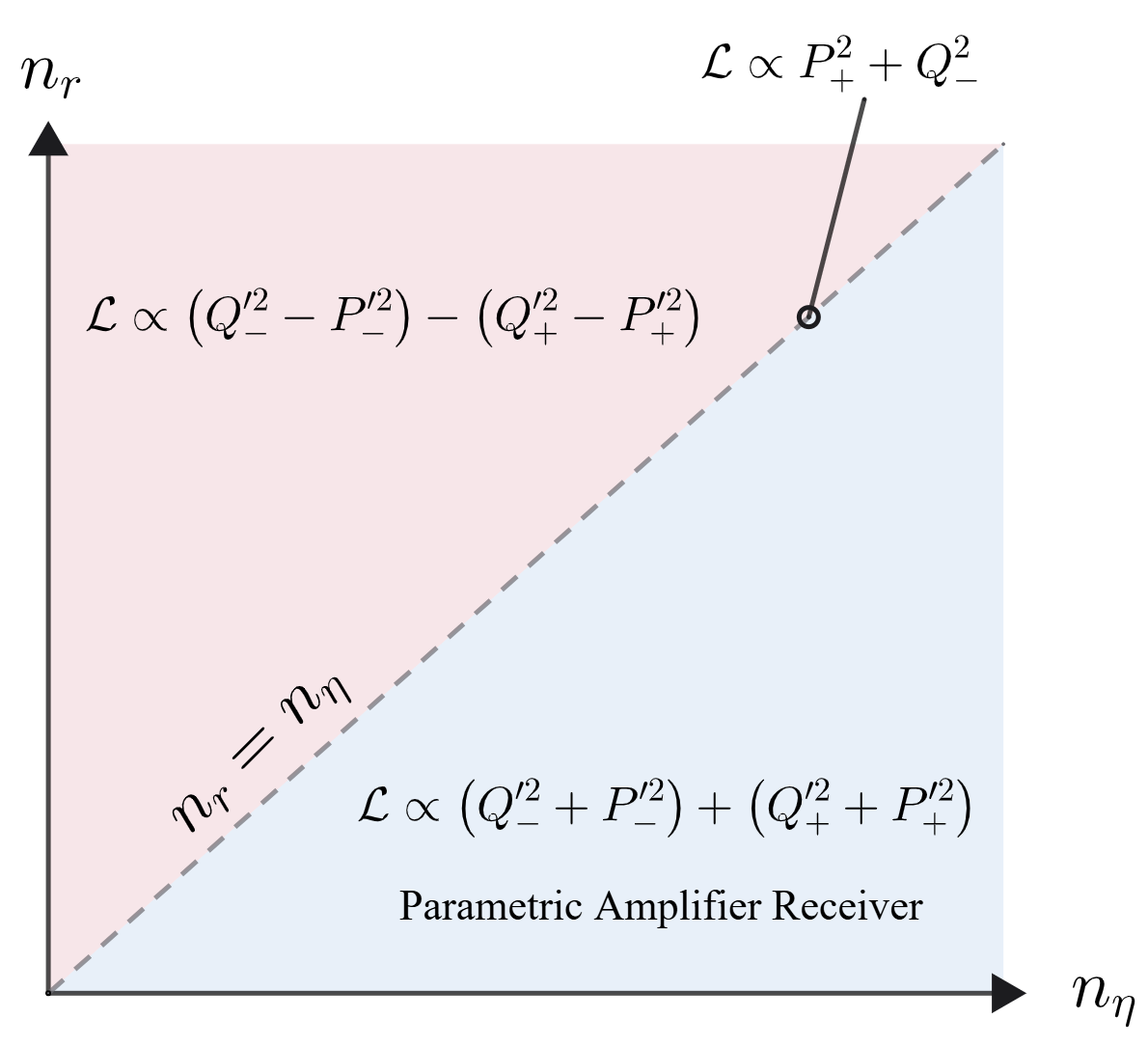}
    \caption{
        \label{fig:SLD_regs}
        SLD regimes: For $n_r < n_\eta$ (blue region),
        the SLD corresponds to
        Eq.~(\ref{SLD:PA_rec}) which can be realized through parametric amplifier receiver.
        At the transition point $n_r = n_\eta$, the SLD reduces to  $Q_{-}^{2}+P_{+}^{2}$, which can be realized by homodyne measurement of $Q_{-},P_{+}.$
        The more challenging regime corresponds to $n_r > n_\eta$ (pink region), in which the SLD is given by Eq.~(\ref{SLD:SQ}).
        This observable in principle requires measuring squeezing generators. 
    }
\end{figure}

The different SLD regimes are illustrated in Fig.~\ref{fig:SLD_regs}, and summarized in the follows:
\begin{enumerate}
    \item \textbf{Transition Point ($n_r=n_\eta$):} The optimal quantum measurement reduces to a non-local double homodyne detection: homodyne measurement of $P_+,$ and $Q_{-}$ (same as Ref. \cite{R37_PhysRevResearch.3.013006}).
    \item \textbf{Reflectivity Dominated Regime ($n_\eta>n_r$):} 
    The optimal detection scheme corresponds to parametric amplifier receiver
    : a pre-processing parametric amplifier followed by number resolving measurement of both modes.
    \item \textbf{Squeezing Dominated Regime ($n_r>n_\eta$):} The optimal observable corresponds to 
    two-mode squeezing generator. 
    While a direct realization of this measurement is challenging, it can be approximated 
    by parametric amplifier receivers \cite{guha2009gaussian} and several Gaussian measurements \cite{R37_PhysRevResearch.3.013006,reichert2023quantum}, as will be discussed in Secs. \ref{susec:homodyne_meas_strategy}, \ref{sec:adaptive_heterodyne_homodyne}.
\end{enumerate}

One possible way to understand this behavior of the SLD is
using the decomposition of the QFI in Eq.~(\ref{eq:qfi_symplectic_contributions}): $\mathcal{I}=\mathcal{I}_\text{vals}+\mathcal{I}_\text{modes},$
where $\mathcal{I}_\text{vals}$ is the information from the symplectic eigenvalues and $\mathcal{I}_\text{modes}$ is the information from the change in the normal modes.
Inserting Eqs.~(\ref{eq:symp_eigenvalues}),(\ref{eq:symp_transformation}) into Eq.~(\ref{eq:qfi_symplectic_contributions}),
we obtain 
that in our case:
\begin{align}
\begin{split}
    \mathcal{I}_\text{modes} &= \frac{(n_r^2+n_r)\tilde{N}_+}{\eta[(1-\eta)\tilde{N}+1]\tilde{N}_-}
    \\
    \mathcal{I}_\text{vals} &= \frac{\tilde{N}_--[(1-\eta)\tilde{N}+1]}{(1-\eta)^2\tilde{N}_-}
\end{split}    
\end{align}
where
$\tilde{N} \equiv n_\beta+n_r+2n_\beta n_r,$
and
$\tilde{N}_\pm \equiv 1+(1-\eta^2)\tilde{N}+(1-\eta)^2(n_r^2+n_r)
+(1\pm\eta)^2(n_\beta^2+n_\beta).$

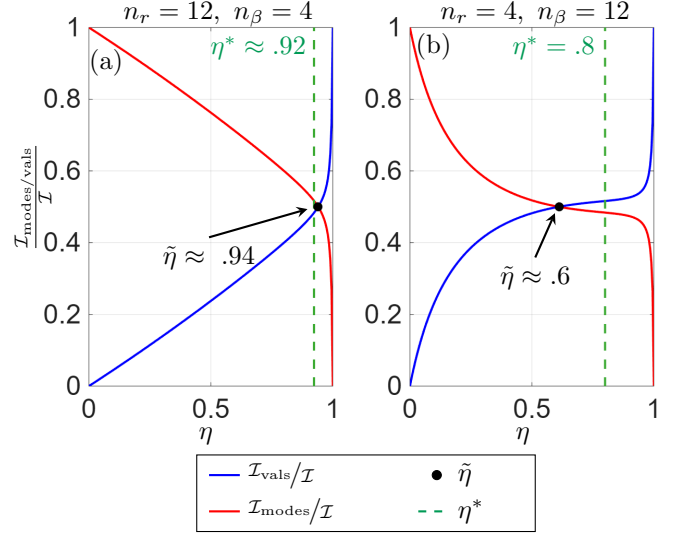
\begin{figure}[H]
    \centering
    \begin{tikzpicture}
        \node[anchor=south west, inner sep=0] (image) at (0,0) {
            \includegraphics[width=3.2in, trim={0mm 0mm 0mm 0mm}, clip]{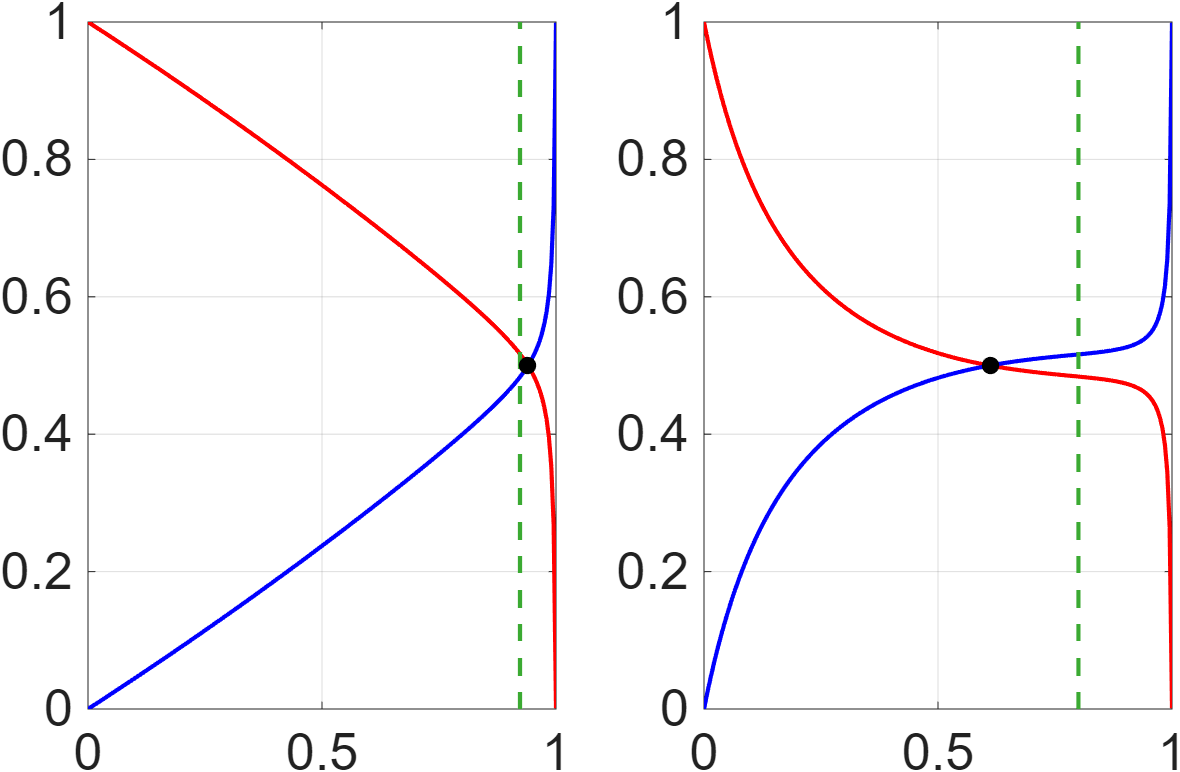}
        };

        \begin{scope}[x={(image.south east)}, y={(image.north west)}]

            \node[above,font=\normalsize] at (.285, .95) {$n_r=12,\ n_\beta=4 $};
            \node[above,font=\normalsize] at (.8, .95) {$n_r=4,\ n_\beta=12 $};

            \node[above,font=\normalsize,black] at (.103, .83) {(a)};
            \node[above,font=\normalsize,black] at (.635, .87) {(b)};
            
            \node[below,font=\normalsize,rotate=90] at (-.06, .55) {$\frac{\mathcal{I}_\text{modes/vals}}{\mathcal{I}}$};

            \node[below,font=\normalsize] at (0.265, 0) {$\eta$};
            \node[below,font=\normalsize] at (0.78, 0) {$\eta$};

            \node[above,font=\normalsize,ForestGreen] at (.83, .87) {$\eta^* = .8$};
            \node[above,font=\normalsize,ForestGreen] at (.35, .87) {$\eta^* \approx .92$};
            
            \draw[->, -stealth, thick,black] (.8, .4) -- (.832, .51);
            \node[below,font=\normalsize,black] at (.8, .4) {$\tilde{\eta} \approx .6$};

            \draw[->, -stealth, thick,black] (.27, .45) -- (.43, .52);
            \node[below,font=\normalsize,black] at (.27, .45) {$\tilde{\eta} \approx\ .94$};

            \draw[fill=white] (.25,-.09) rectangle (.8,-.28);
            
                \node[right,font=\normalsize] at (.32,-.14) {$\nicefrac{\mathcal{I}_{\text{vals}}}{\mathcal{I}}$};
                \draw[-,blue,thick] (.27,-.14) -- (.32,-.14);

                \node[right,font=\normalsize] at (.32,-.23) {$\nicefrac{\mathcal{I}_{\text{modes}}}{\mathcal{I}}$};
                \draw[-,red,thick] (.27,-.23) -- (.32,-.23);

                \node[right,font=\normalsize] at (.66,-.14) {$\tilde{\eta}$};
                \draw[fill=black] (.64,-.14) circle (.6mm);

                \node[right,font=\normalsize] at (.66,-.23) {$\eta^*$};
                \draw[dashed,ForestGreen,thick] (.61,-.23) -- (.66,-.23);

        \end{scope}
    \end{tikzpicture}
    \caption{
        \label{fig:W_decomp}  
        The ratios $\nicefrac{\mathcal{I}_{\text{vals}}}{\mathcal{I}}$ and $\nicefrac{\mathcal{I}_{\text{modes}}}{\mathcal{I}}$ as a function of $\eta$ for: (a) $n_r =12 $ and $n_\beta=4$ (b) $n_r =4 $ and $n_\beta=12$. The green dashed line indicated by $\eta^*$ represent the $n_\eta=n_r$ line, and the black dot is the information equilibrium point in which $\mathcal{I}_{\text{vals}}=\mathcal{I}_{\text{modes}}$ indicated by $\tilde{\eta}$.
    }
\end{figure}

It can be thus observed that for $\eta \rightarrow 0,$ $\mathcal{I}_\text{modes} \gg \mathcal{I}_\text{vals},$
hence in this limit the information comes almost entirely from the symplectic transformation.
In the other limit of $\eta \rightarrow 1,$
we have $\mathcal{I}_\text{modes} \gg \mathcal{I}_\text{vals},$
such that almost all of the information comes from the symplectic eigenvalues.
This is illustrated in Fig. \ref{fig:W_decomp}.

There is therefore a transition from the small $\eta$ regime in which the contribution from the symplectic transformation is dominant to a large $\eta$ regime in which the contribution from the symplectic eigenvalues is dominant.
Although this transition point is not identical to that of the SLD, as illustrated in Fig.~\ref{fig:W_decomp}, it provides physical intuition for the different SLD regimes.
The information encoded in the symplectic eigenvalues is obtained via parametric amplifier receiver (by performing the inverse symplectic transformation followed by number resolving measurement). Hence in the large $\eta$ regime the optimal measurement corresponds to parametric amplifier receiver. Whereas the information from the symplectic transformation is obtained via squeezing generator measurements, hence in the limit of small $\eta$ this is the optimal measurement.

In the next subsection we further explore the performance of non-local homodyne measurements.
We observed that a particular non-local homodyne strategy is optimal at $\eta=\eta^*$. 
This raises the question of whether this strategy remains near-optimal beyond this specific point, 
and whether we can obtain further improvement by considering optimized non-local homodyne measurements. These questions are addressed in the next subsection.

\subsection{Non-Local Homodyne Measurement Strategy}
\label{susec:homodyne_meas_strategy}
Let us explore the performance of non-local homodyne strategy.

We first study the performance of $P_+,Q_-$ measurement, which was shown to be optimal at the transition point of $n_\eta=n_r$.
A more general class of non-local measurements is then considered
and optimized. 

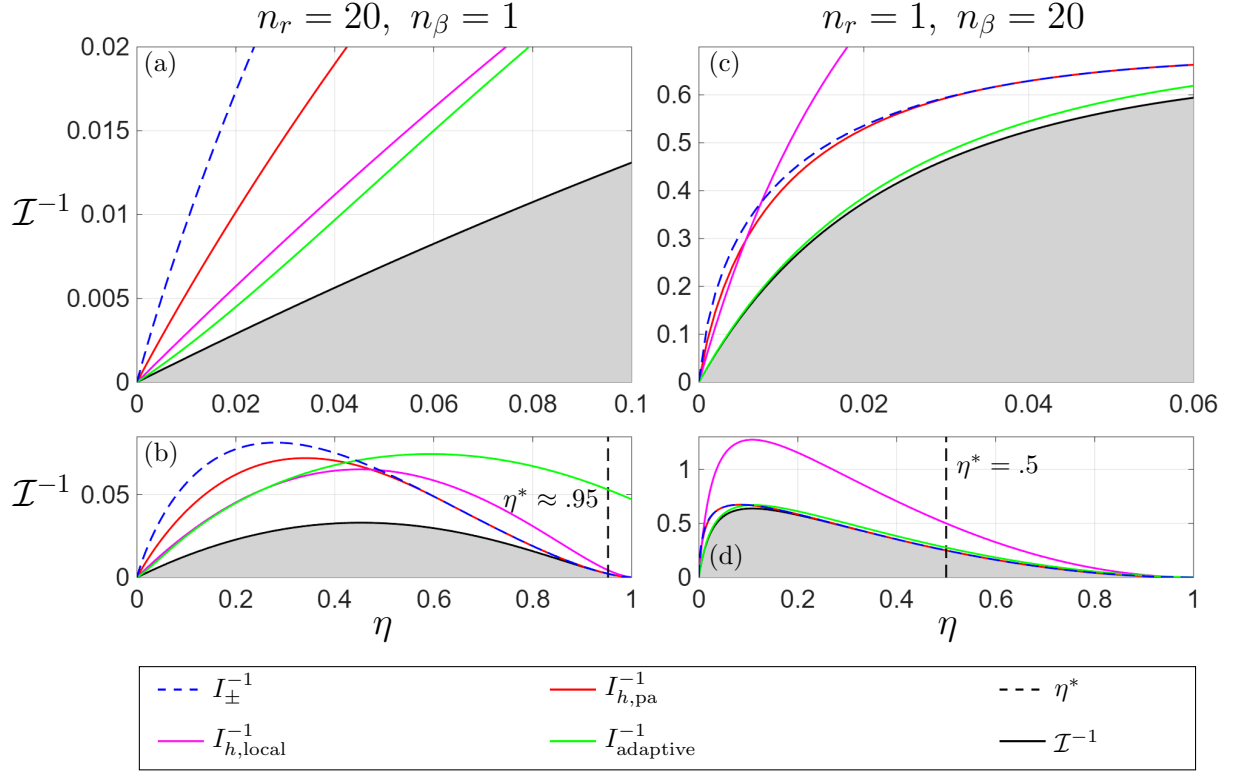
\begin{figure*}[t]
    \centering
    \begin{tikzpicture}
        \node[anchor=south west, inner sep=0] (image) at (0,0) {
            \includegraphics[width=6in, trim={0mm 0mm 0mm 0mm}, clip]{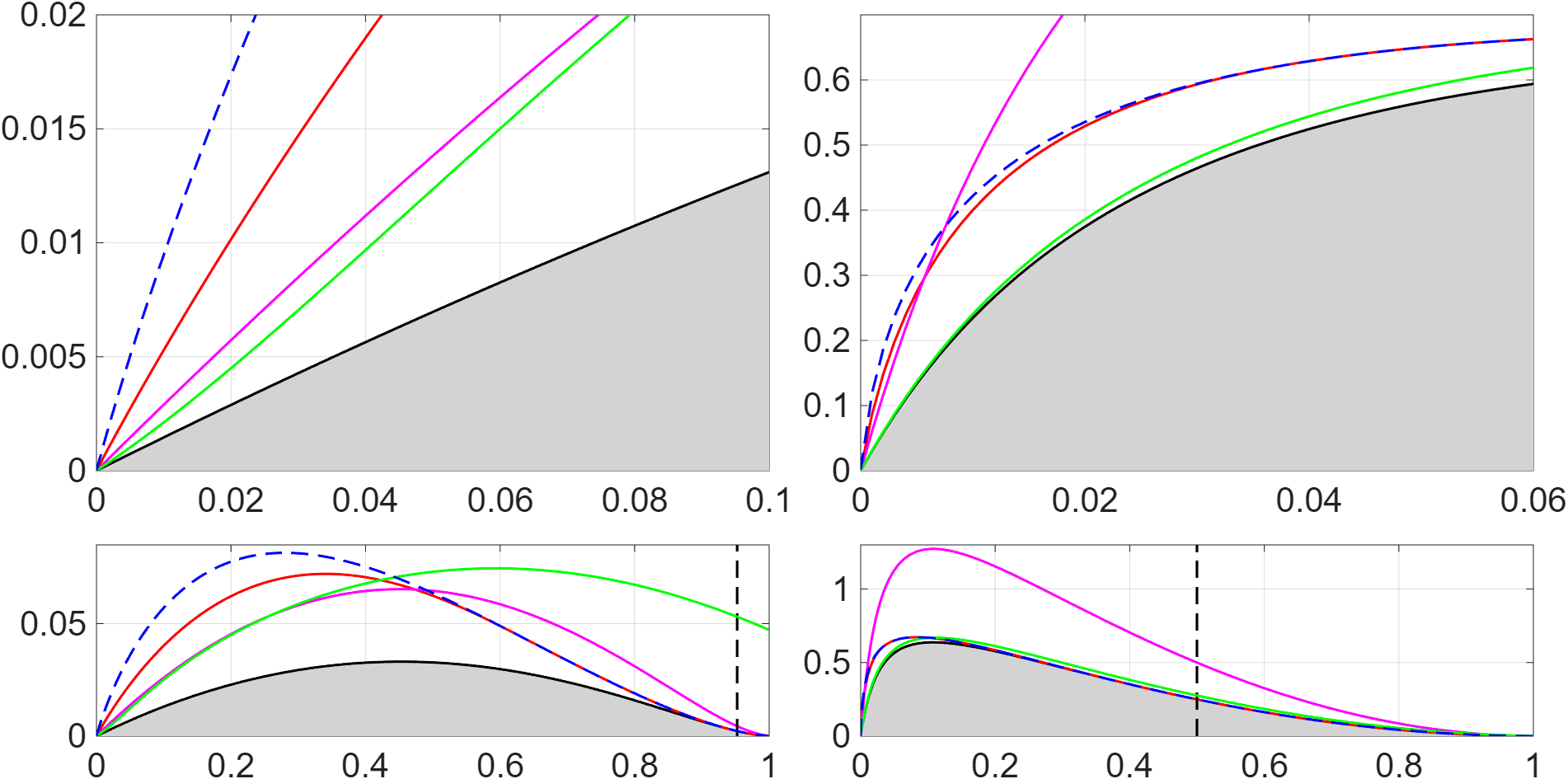}
        };

        \begin{scope}[x={(image.south east)}, y={(image.north west)}]

           \node[below,font=\Large,rotate=0] at (-.02, .74) {$\mathcal{I}^{-1}$};
            \node[below,font=\Large,rotate=0] at (-.02, .25) {$\mathcal{I}^{-1}$};
            \node[below,font=\Large] at (0.275, 0) {$\eta$};
            \node[below,font=\Large] at (0.765, 0) {$\eta$};

            \node[right,font=\normalsize] at (.06, .95) {(a)};
            \node[right,font=\normalsize] at (.06, .27) {(b)};
            \node[right,font=\normalsize] at (.55, .95) {(c)};
            \node[right,font=\normalsize] at (.55, .09) {(d)};

            \node[right,font=\normalsize] at (.765, .25) {$\eta^*=.5$};
            \node[left,font=\normalsize] at (.47, .19) {$\eta^*\approx.95$};

            \node[above,font=\Large] at (.28, .98) {$n_r=20,\ n_\beta=1 $};
            \node[above,font=\Large] at (.77, .98) {$n_r=1,\ n_\beta=20 $};

            \draw[fill=white] (.064,-.1) rectangle (.98,-.28);
            
                \node[right,font=\normalsize] at (.12,-.14) {$I_\pm^{-1}$};
                \draw[dashed,blue,thick] (.08,-.14) -- (.12,-.14);

                \node[right,font=\normalsize] at (.12,-.23) {$I_{h,\text{local}}^{-1}$};
                \draw[-,draw={rgb:RubineRed,.9;VioletRed,1},thick] (.08,-.23) -- (.12,-.23);

                \node[right,font=\normalsize] at (.46,-.14) {$I_{h,\text{pa}}^{-1}$};
                \draw[-,red,thick] (.42,-.14) -- (.46,-.14);

                \node[right,font=\normalsize] at (.46,-.23) {$I_{\text{adaptive}}^{-1}$};
                \draw[-,green,thick] (.42,-.23) -- (.46,-.23);

                \node[right,font=\normalsize] at (.85,-.14) {$\eta^*$};
                \draw[dashed,black,thick] (.81,-.14) -- (.85,-.14);
                
                \node[right,font=\normalsize] at (.85,-.23) {$\mathcal{I}^{-1}$};
                \draw[-,black,thick] (.81,-.23) -- (.85,-.23);

        \end{scope}
    \end{tikzpicture}
    \caption{
    CRB of different Gaussian detection schemes: local homodyne ($I_{h,\text{local}}$), non-local homodyne ($I_\pm$), adaptive heterodyne-homodyne ($\mathcal{I}_{\text{adaptive}}$) and PA-assisted non-local homodyne ($I_{h,\text{pa}}$), 
    compared to the QFI ($\mathcal{I}$). The transition point between the different SLD regimes, $\eta=\eta^{*},$ is marked with a dashed black line. (a-b) $n_r=20, n_\beta=1.$ In this regime the non-local homodyne saturates the QFI only in a narrow neighborhood of the transition point $\eta=\eta^{*}$. (c-d) $n_r=1, n_\beta=20$. In this regime the non-local homodyne saturates the QFI over a broader range of $\eta.$ For small values of $\eta,$ the adaptive heterodyne-homodyne strategy saturates the QFI and outperforms non-local homodyne, as shown on the top panel. }
    \label{fig:I_opt}
\end{figure*}

Consider the $P_+,Q_-$ measurement. The FI obtained with this measurement is
\begin{align}
I_{\pm}=\frac{\left(n_{\beta}-n_{r}+\frac{\sqrt{n_r^2+n_{r}}}{\sqrt{\eta}}\right)^{2}}{\Big[1+n_{\beta}\left(1-\eta\right)+n_{r}\left(1+\eta\right)-2\sqrt{\eta}\sqrt{n_r^2+n_{r}}\Big]^{2}}.    
\end{align}
The behavior of this FI for different parameter regimes is shown in Fig.~\ref{fig:I_opt}.
While this measurement is always optimal at $n_\eta=n_r,$ it is also nearly optimal for a much wider range of parameters.
The optimality of this measurement depends on $n_\beta$: in the large $n_\beta$ regime it is nearly optimal.
It can be seen that for $n_\beta \gg \frac{n_r}{\sqrt{\eta}},\frac{1}{\sqrt{\eta}},$
this FI converges to $1/\left( 1-\eta \right)^2,$
which is also the QFI in this limit.
On the other hand,
in the large squeezing regime
this measurement is not optimal ; In the limit of $n_r \gg \frac{n_{\beta}}{1-\sqrt{\eta}}, 1,$
this FI converges to $\frac{1}{\eta \left(1-\sqrt{\eta} \right)^{2}}$, while the QFI grows linearly with $n_r$ (Eq.~\ref{eq:qfi_large_nr}) and can be thus much larger.

Let us now consider a more general class of non-local homodyne measurements: quadratures that can be measured via pre-processing parametric amplifier transformation. 
A parametric amplifier transformation with squeezing coefficient $r_{\text{pre}}$ applies the following transformation to the quadratures:
\begin{align}
\begin{split}
&S(r_{\text{pre}})\left[\begin{array}{c}
Q_{i}\\
P_{i}\\
Q_{s}\\
P_{s}
\end{array}\right]=\left[\begin{array}{c}
Q'_{i}\left(r_{\text{pre}}\right)\\
P'_{i}\left(r_{\text{pre}}\right)\\
Q'_{s}\left(r_{\text{pre}}\right)\\
P'_{s}\left(r_{\text{pre}}\right)
\end{array}\right]=\\
&\left[\begin{array}{c}
\cosh\left(r_{\text{pre}}\right)Q_{i}+\sinh\left(r_{\text{pre}}\right)Q_{s}\\
\cosh\left(r_{\text{pre}}\right)P_{i}-\sinh\left(r_{\text{pre}}\right)P_{s}\\
\cosh\left(r_{\text{pre}}\right)Q_{s}+\sinh\left(r_{\text{pre}}\right)Q_{i}\\
\cosh\left(r_{\text{pre}}\right)P_{s}-\sinh\left(r_{\text{pre}}\right)P_{i}
\end{array}\right].
\label{eq:parametric_amplifier_quadtratures}
\end{split}
\end{align}
Optimizing over this class of measurements means optimizing the FI over $r_{\text{pre}}$ and over the choice of two commuting quadratures from the set of Eq.~(\ref{eq:parametric_amplifier_quadtratures}).
Note that there are four possible pairs of commuting quadratures: $\left\{ Q'_{i}\left(r_{\text{pre}}\right),Q'_{s}\left(r_{\text{pre}}\right)\right\}$ , $\left\{ P'_{i}\left(r_{\text{pre}}\right),P'_{s}\left(r_{\text{pre}}\right)\right\}$ , $\left\{ Q'_{i}\left(r_{\text{pre}}\right),P'_{s}\left(r_{\text{pre}}\right)\right\}$ , $\left\{ P'_{i}\left(r_{\text{pre}}\right),Q'_{s}\left(r_{\text{pre}}\right)\right\}$. It can be seen that the FI with the first two pairs: 
$\left\{ Q'_{i}\left(r_{\text{pre}}\right),Q'_{s}\left(r_{\text{pre}}\right)\right\}$ , $\left\{ P'_{i}\left(r_{\text{pre}}\right),P'_{s}\left(r_{\text{pre}}\right)\right\}$
is exactly the same as with optimal local homodyne. 
It can be then further observed that optimizing $\left\{ Q'_{i}\left(r_{\text{pre}}\right),P'_{s}\left(r_{\text{pre}}\right)\right\}$ over $r_{\text{pre}}$
is equivalent to optimizing $\left\{ P'_{i}\left(r_{\text{pre}}\right),Q'_{s}\left(r_{\text{pre}}\right)\right\},$
since these pairs of quadratures can be obtained from one another by a local rotation in the phase space and $r_{\text{pre}} \mapsto -r_{\text{pre}}.$
Hence, optimizing over this class of measurements can be reduced to optimizing the FI over $r_{\text{pre}}$ given a measurement of $\left\{ Q'_{i}\left(r_{\text{pre}}\right),P'_{s}\left(r_{\text{pre}}\right)\right\} $.

Note that in the limit of $r_\text{pre} \rightarrow -\infty$
these quadratures correspond to $\left\{ Q_{-}, P_+ \right \}$,
hence this parametric-amplifier assisted measurement generalizes the  $\left\{ Q_{-}, P_+ \right \}$ detection scheme.

The covariance matrix given a measurement of $\left\{ Q'_{i}\left(r_{\text{pre}}\right),P'_{s}\left(r_{\text{pre}}\right)\right\} $ is :
\begin{align}
    \Sigma_{Q'_{i}\left(r_{\text{pre}}\right),P'_{s}\left(r_{\text{pre}}\right)}=
    \begin{bmatrix}
        n_d+\nicefrac{1}{2} & 0 \\[6pt]
         0 & n_d+\nicefrac{1}{2}-\delta_n\eta
    \end{bmatrix},
    \label{eq:Covqp}
\end{align}
where the effective photon number $n_d$ is given by:
\begin{equation}
\begin{aligned}
    n_d &= 
    \frac{1}{2}\left[ \cosh(2r_\text{pre})(n_r+n_\beta)-\delta_n \right]
    \\&+
    \sinh(2r_\text{pre})\sqrt{\eta(n_r^2+n_r)} + \sinh^2(r_\text{pre}).
\end{aligned}
\end{equation}
The FI of these quadratures is optimized numerically over $r_\text{pre}.$
The optimal FI $I_{\text{h,pa}}$ is shown in Fig. \ref{fig:I_opt}, where it is also compared to the QFI, the local homodyne FI, and the $P_+,Q_-$ measurement. In Appendix~\ref{app:optimization} we show that for most of the parameters space, the optimal PA-assisted measurement corresponds to the measurement of $P_+,Q_-.$

Fig.~\ref{fig:I_opt}(a-b) characterizes the performance in the regime of $n_r \gg n_\beta$. In this regime, optimized non-local homodyne (red) saturates the QFI only in a neighborhood of $\eta=\eta^{*}$, which in this illustration corresponds to $\eta \gtrsim 0.95$. For smaller $\eta$ optimized non-local homodyne becomes much less efficient and even underperforms local homodyne detection (magenta line). We observe a distinct crossover: for low reflectivities (in this illustration $\eta \lesssim 0.56$), the local strategy performs better, whereas for higher reflectivities the non-local strategy becomes advantageous. 

In the regime of $n_r \ll n_\beta$, depicted in Fig.~\ref{fig:I_opt}(c-d), optimized non-local strategy (red) saturates the quantum limit at a significantly extended range of $\eta$. In this illustration, the optimized non-local strategy saturates the QFI for $\eta \gtrsim 0.2$. While in this case there is also a crossover between local and non-local homodyne strategies, this transition occurs at a much lower $\eta$. In this illustration the crossover is at $\eta \approx 4.1 \times 10^{-3}$, rendering the optimized non-local strategy the preferred method for the vast majority of the parameter space. 

This analysis shows that optimized non-local homodyne detection saturates the QFI in the regime of $n_{\beta}\gg n_{r}/\sqrt{\eta},1$. When $\eta \ll 1$ or $n_r$ is large, these conditions break down and optimized non-local homodyne does not attain the QFI. We show in the next subsection that in these cases a different Gaussian detection scheme, an adaptive heterodyne-homodyne scheme, can be more efficient.

\subsection{Adaptive heterodyne-homodyne measurements}
\label{sec:adaptive_heterodyne_homodyne}
Following Refs.~\cite{shi2023fulfilling,reichert2023quantum} we consider an adaptive heterodyne-homodyne measurement. First, a heterodyne measurement is performed on one of the modes (the idler, the signal or a combination thereof) resulting in an outcome of $\alpha\in\mathbb{C}$. A quadrature measurement is then performed on the remaining mode, with the measured quadrature chosen adaptively based on $\alpha$. We focus on the following protocol: heterodyne of the signal mode followed by homodyne of the idler, and show that it provides advantage mainly in the quantum illumination limit. 

Consider a heterodyne measurement of the signal mode followed by homodyne measurement of the idler. Given an outcome of $\alpha\in\mathbb{C}$ of the heterodyne measurement, the idler mode is in a Gaussian state given by the mean vector $\boldsymbol{\mu}=\frac{1}{\sqrt{2}}\frac{\sqrt{\eta}\sinh\left(2r\right)}{\eta n_{r}+\left(1-\eta\right)n_{\beta}+1}\left[\text{Re}\left(\alpha\right),-\text{Im}\left(\alpha\right)\right]^{t},$ and the covariance matrix is $V_i=\frac{1}{2}\left[\cosh\left(2r\right)-\frac{1}{2}\frac{\eta\sinh\left(2r\right)^{2}}{\eta n_{r}+\left(1-\eta\right) n_{\beta}+1}\right]\mathbb{I}_{2}$. The optimal homodyne measurement of this state is along its mean vector, and the resulting FI is then (see Appendix ~\ref{app:adaptive_heterodyne_homodyne} for details):
\begin{align}
\begin{split}
&I_{\text{adaptive}}=\frac{1}{\left(1+n_{\beta}\left(1-\eta\right)+n_{r}\eta\right)^{2}}\big[ \left(n_{\beta}-n_{r}\right)^{2}+\\
&\frac{\bigl(1+n_{\beta}+n_{\beta}\eta-n_{r}\eta\bigr)^{2}n_{r}\left(n_{r}+1\right)}{\eta\left(1+n_{r}(2-\eta)+n_{\beta}\left(1+2n_{r}\right)\left(1-\eta\right)\right)}+\\
&\frac{2\left(1+n_{\beta}\right)^{2}n_{r}^{2}\left(n_{r}+1\right)^{2}}{\left(1+n_{r}(2-\eta)+n_{\beta}\left(1+2n_{r}\right)\left(1-\eta\right)\right)^{2}} \big].
\end{split}    
\end{align}
In the limit of $\eta \ll 1$ this FI reduces to $I_{\text{adaptive}}\approx\frac{n_{r}\left(n_{r}+1\right)}{\eta\left(1+2n_{r}+n_{\beta}\left(1+2n_{r}\right)\right)}.$
In the quantum illumination regime of $n_r \ll1\ll n_{\beta}$ it becomes $I_{\text{adaptive}}\approx\frac{n_{r}}{\eta n_{\beta}}$ which coincides with the QFI in this limit.
This is illustrated in Fig.~\ref{fig:I_opt}. This figure also presents a comparison between the adaptive hetero-homodyne protocol and non-local and local homodyne schemes.

We can also consider a heterodyne of the idler mode followed by adaptive homodyne of the signal, this scheme however yields the same FI as the classical coherent state (Eq. ~\ref{eq:FI_homodyne_coherent}).
To see this, note that the heterodyne measurement of the idler commutes with the thermal channel of the signal.
Therefore this strategy is equivalent to performing heterodyne of the idler before the encoding thermal channel of the signal mode.
The state of the signal mode after an idler heterodyne measurement with outcome $\alpha$ corresponds to a coherent state with a mean photon number
$\frac{n_{r}}{n_{r}+1}|\alpha|^{2}$.
Averaging over the outcome of the heterodyne measurement, $\alpha,$ we obtain the coherent state FI of Eq.~(\ref{eq:FI_homodyne_coherent}) (see Appendix.~\ref{app:adaptive_heterodyne_homodyne} for details). This can be viewed as another simple proof for the advantage of the TMSV over coherent states: there is a measurement of the TMSV that retrieves the coherent state QFI.

This scheme thus simply recovers the coherent state performance, and does not present any advantage compared to the classical case.

\section{Conclusions and outlook \label{sec:V}}
In this work, we studied 
the 
quantum
precision limits for 
target reflectivity
estimation using a TMSV probe
under various classes of detection schemes.

Restricting the measurements to local homodyne readout severely limits the performance of the TMSV, such that in certain regimes even the classical coherent state strategy outperforms it.

Using the SLD, we identified measurements that saturate the QFI and found a transition in the optimal measurement.
In the reflectivity dominated regime ($n_\eta>n_r$),
parametric amplifier receivers are optimal.
By contrast, in the
squeezing dominated regime, ($n_r>n_\eta$), 
such receivers do not necessarily saturate the QFI and the two-mode squeezing generators become the optimal observables. Of particular interest is the transition point, ($n_r=n_\eta$), in which a non-local homodyne measurement 
is  optimal.  We then studied two Gaussian measurement schemes: an optimized PA-assisted non-local homodyne strategy and an adaptive heterodyne-homodyne scheme. 
In high-noise environments, 
these approaches provide complementary advantages. The non-local homodyne strategy approaches the QFI for high reflectivity, whereas the adaptive scheme saturates it for very low reflectivity. 
These results demonstrate that 
Gaussian measurements can saturate, or closely approach, the QFI over a wide range of the problem parameters.

Several open questions are left for future work. A natural open question is whether there exist Gaussian measurements that closely approach the QFI for any given parameters of the thermal loss channel.
In particular, both the non-local homodyne strategy and the adaptive heterodyne-homodyne schemes can be further optimized.
It would be interesting to develop a more general characterization of optimal measurements for estimating arbitrary parameters of the thermal loss channel, beyond the loss parameter considered here.

\textbf{Acknowledgments--} The authors are thankful to Roberto Di Candia, Matteo Fadel, and James W. Gardner for useful discussions.
TG acknowledges funding provided by the Quantum Science and Technology early career fellowship of the Israel Council for Higher Education
and ISF Grant No. 3302/25. NK acknowledges support of EU project OpenSuperQPlus and ISF Grant No. 2128/24. 

\appendix
\section{Analysis of $\varepsilon^-$\label{apx:eps}}
Assuming non-negative $r$, the explicit expression for $\varepsilon^-$ is given by:
\begin{equation}
    \varepsilon^-(n_\eta,n_r) = \frac{N+2n_\eta n_r+1 - \sqrt{\delta_n^2+4(n_r^2+n_r)(n_\eta^2+n_\eta)}}{2(n_\eta+1)}
\end{equation}  
where $N = n_r+n_\eta+n_\beta$.  

In the limit of perfect reflectivity $\eta \to 1$ (where $n_\eta \to \infty$), the eigenvalue expression simplifies to:
\begin{equation}
    \lim_{n_\eta\to\infty}\varepsilon^- = \left(n_r+\frac{1}{2}\right)-\sqrt{n_r^2+n_r}
\end{equation}
In this regime, $\varepsilon^- \leq \frac{1}{2}$, indicating that the system is entangled for all $n_r$. Furthermore, $\varepsilon^-$ becomes invariant with respect to $n_\beta$. This result is physically consistent: because the target acts as a perfect mirror, the reflected signal remains isolated from environmental noise. Consequently, as indicated by the dashed blue curve (Fig.~\ref{fig:ent_cond}), $\varepsilon^- \to 0$ as $n_r$ increases. This implies that the non-zero asymptotic limit observed in the general case (solid blue curves) arises solely from thermalization noise. Theoretically, with a perfect mirror, entanglement persists regardless of the environmental noise level $n_\beta$.

At the opposite extreme, where $\eta \to 0$ (dashed red curve), we find:
\begin{equation}
    \lim_{n_\eta\to0}\varepsilon^- = \frac{1}{2}\Big[(n_r+n_\beta)-|\delta_n|+1\Big]
\end{equation}
In this scenario, the probe beam is completely lost. This yields $\varepsilon^- \geq \frac{1}{2}$ for all parameters. Consequently, entanglement cannot be sustained in this limit.
\\ \\
Next, we examine the asymptotic behavior of $\varepsilon^-$. The evaluation of limits must be handled carefully to reflect the experimental hierarchy. Since the probe energy $n_r$ is a controllable parameter while the target reflectivity $n_\eta$ is an unknown system property, we analyze the limit with respect to $n_r$.

We find that:
\begin{equation}
    \lim_{n_r\to\infty}\partial_{n_r}\varepsilon^-=0
\end{equation}  
This indicates that at sufficiently high $n_r$, further increases in squeezing yield negligible improvements in the entanglement level, a saturation effect visible in Fig.~\ref{fig:ent_cond}. This behavior provides a crucial guideline for experimental resource allocation: in a controlled setting, one must determine whether it is more advantageous to invest in lowering the environmental temperature or in increasing the parametric amplifier power.

\section{Optimal Local Homodyne Measurements}
\label{app:optimal_local_homodyne}
In this appendix, we show that the optimal local homodyne strategy consists of measuring identical quadratures on both modes (e.g., $Q_s$ and $Q_i$).

Due to the rotational symmetry of the TMSV state, we can parameterize the general local homodyne measurement as follows: we fix the measurement on the signal mode to be $Q_s$ and vary the measurement on the idler mode as a linear combination of quadratures: 
$\sqrt{p}Q_{i} + \sqrt{1-p}P_{i}$. Where $0 \leq p \leq 1$. Note that the sign of the $P_i$ component does not affect the covariance matrix or the FI, allowing us to restrict our analysis to the positive branch.

The FI, $I_{h,\text{local}}(p),$ for this configuration is given by:
\begin{align}
\begin{split}
&I_{h,\text{local}}(p)=\frac{1}{\eta\Bigl(1+2\tilde{N}(1-\eta)+4\eta n_{r}\left(1+n_{r}\right)\left(1-p\right)\Bigr)^{2}}\\
&\times \Bigg[ (1+2n_{\beta})n_{r}(1+n_{r})(1+2n_{r})p \\
&\quad +2\bigl[n_{\beta}(1+2n_{r})-n_{r}(1+2n_{r}) +2n_{r}(1+n_{r})p\bigr] \\
&\quad \times \bigl[\tilde{N}+n_{r}(n_{r}+1)(p-2)\bigr]\eta\Bigg].
\end{split}
\end{align}

A direct calculation of the second derivative confirms that the FI is convex with respect to $p$:
\begin{equation}
    \frac{\partial^{2}I_{h,\text{local}}}{\partial p^2} \geq 0 \quad \forall \ \{n_r, n_\beta, \eta, p\}.
\end{equation}

Therefore, its maximum must occur at one of the boundaries of the domain: $p=0$ or $p=1$. Evaluating the function at these endpoints shows that the maximum is  achieved at $p=1$. This corresponds to measuring $Q_i$ (identical to the signal quadrature $Q_s$), yielding the FI presented in Eq.~(\ref{eq:HFI}).

\section{Bound on the Performance Ratio ($\mathcal{R} \leq 1$)\label{apx:Rgeq1}}
By comparing Eq.~(\ref{eq:QFI}) and Eq.~(\ref{eq:QFI_coherent}), we determine the condition in which the quantum advantage vanishes (i.e., $\mathcal{I}=\mathcal{I}_{coh}$). This equality occurs when $n_r = n_{r,eq}$:
\begin{equation}
    n_{r,eq} = -\frac{n_\beta^2\eta(1-\eta)(2+\eta) + n_\beta[\eta(1-\eta)+(1+\eta)]+\eta}{\eta(1-\eta)[n_\beta(1-\eta)+1]}.
\end{equation}
Given the physical constraint $n_{r,eq}\geq0$  
, the only permissible solution to the expression above corresponds to the vacuum limit $n_{r,eq}=0$. Solving for the thermal noise threshold $n_{\beta,eq}$ that satisfies $n_{r,eq}(n_{\beta,eq}^{\pm})=0$ yields:
\begin{equation}
    n_{\beta,eq}^{\pm} = -\frac{(1+2\eta-\eta^2)\pm\sqrt{1-2\eta^2+5\eta^4}}{2(1-\eta)(1+\eta)^2}.
\end{equation}
We observe that the $n_{\beta,eq}^{+}$ solution is strictly negative and thus physically invalid. The $n_{\beta,eq}^{-}$ solution is valid only at the boundaries $\eta=0$ and $\eta=1$, with the limiting behaviors:
\begin{equation}
    \begin{aligned}
        &\lim_{\eta=0} n_{\beta,eq}^{-} = 0,\\
        &\lim_{\eta=1} n_{\beta,eq}^{-} = \infty.
    \end{aligned}
\end{equation}
Consequently, the equality $\mathcal{I}=\mathcal{I}_{coh}$ holds only under one of following conditions:
\begin{enumerate}
    \item $n_r=0$ 
    \item $\eta=0$ and $n_\beta=0$ 
    \item $\eta\to1$ and $n_\beta\to\infty$
\end{enumerate}
Because $\mathcal{R}$ is a continuous function of the system parameters, determining the sign of $\mathcal{R}$ at a single point allows us to generalize the behavior for the entire parameter space. Computing the ratio for the test values $n_r=n_\beta=1$ and $\eta=0.5$, we find $\mathcal{R}=\nicefrac{7}{12}<1$. We therefore conclude that $\mathcal{R} \leq 1$ for all physical values of $n_\beta, n_r$, and $\eta$, confirming the general superiority of the TMSV probe.

\section{Optimal Parametric Amplifier Receivers}
\label{app:optimal_parametric_amplifier}

In this appendix we study the performance of optimized parametric amplifier receivers.
While these receivers coincide with the SLD for $\eta>\eta^{*}$ and thus optimal in this regime, it is unclear how close can they get to the QFI in the $\eta \leq \eta^{*}$ regime. We numerically analyze the performance of these receivers in this regime. 

The following measurement scheme is considered: a parametric amplifier with a squeezing coefficient $r' \in \mathbb{R}$ is applied on the state (after the thermal-loss channel), followed by a number resolving measurement of the two modes.
To calculate the FI with this measurement let us first derive the distribution of the outcomes $\left\{ p\left(m,n\right)\right\} _{m,n\in\mathbb{N}}$.
The state after the thermal-loss channel is the zero-mean Gaussian state given by the covariance matrix of Eq.~(\ref{eq:Cov}). After applying a parametric amplifier with squeezing coefficient $r'$ the state remains a zero-mean Gaussian state and its covariance matrix is now given by
$\Sigma=1/2\left(\begin{array}{cc}
A'\hat{\mathbb{I}} & C'\hat{\sigma}_{z}\\
C'\hat{\sigma}_{z} & B'\hat{\mathbb{I}}
\end{array}\right),$
where
\begin{align}
\begin{split}
&A'=2\sinh\left(2r'\right)\sqrt{\eta n_{r}(n_{r}+1)}+\cosh\left(r'\right)^{2}(2n_{r}+1)\\
&+
\sinh\left(r'\right)^{2}\left[2\eta n_{r}+2n_{\beta}(1-\eta)+1\right],\\
& B'=2\sinh\left(2r'\right)\sqrt{\eta n_{r}(n_{r}+1)}+\sinh\left(r'\right)^{2}(2n_{r}+1)\\
&+
\cosh\left(r'\right)^{2}\left[2\eta n_{r}+2n_{b}(1-\eta)+1\right],\\
& C'=2\cosh\left(2r'\right)\sqrt{\eta n_{r}\left(n_{r}+1\right)}\\
&+\sinh\left(2r'\right)\left(1+n_{r}\left(1+\eta\right)+n_{\beta}\left(1-\eta\right)\right).
\end{split}
\end{align}
The mean photon number of the idler and signal modes is then $N_{i}=\frac{A'-1}{2}$, $N_{s}=\frac{B'-1}{2}$.

\begin{figure}[H]
    \centering
    \begin{tikzpicture}
        \node[anchor=south west, inner sep=0] (image) at (0,0) {
            \includegraphics[width=3.2in, trim={0mm 0mm 0mm 0mm}, clip]{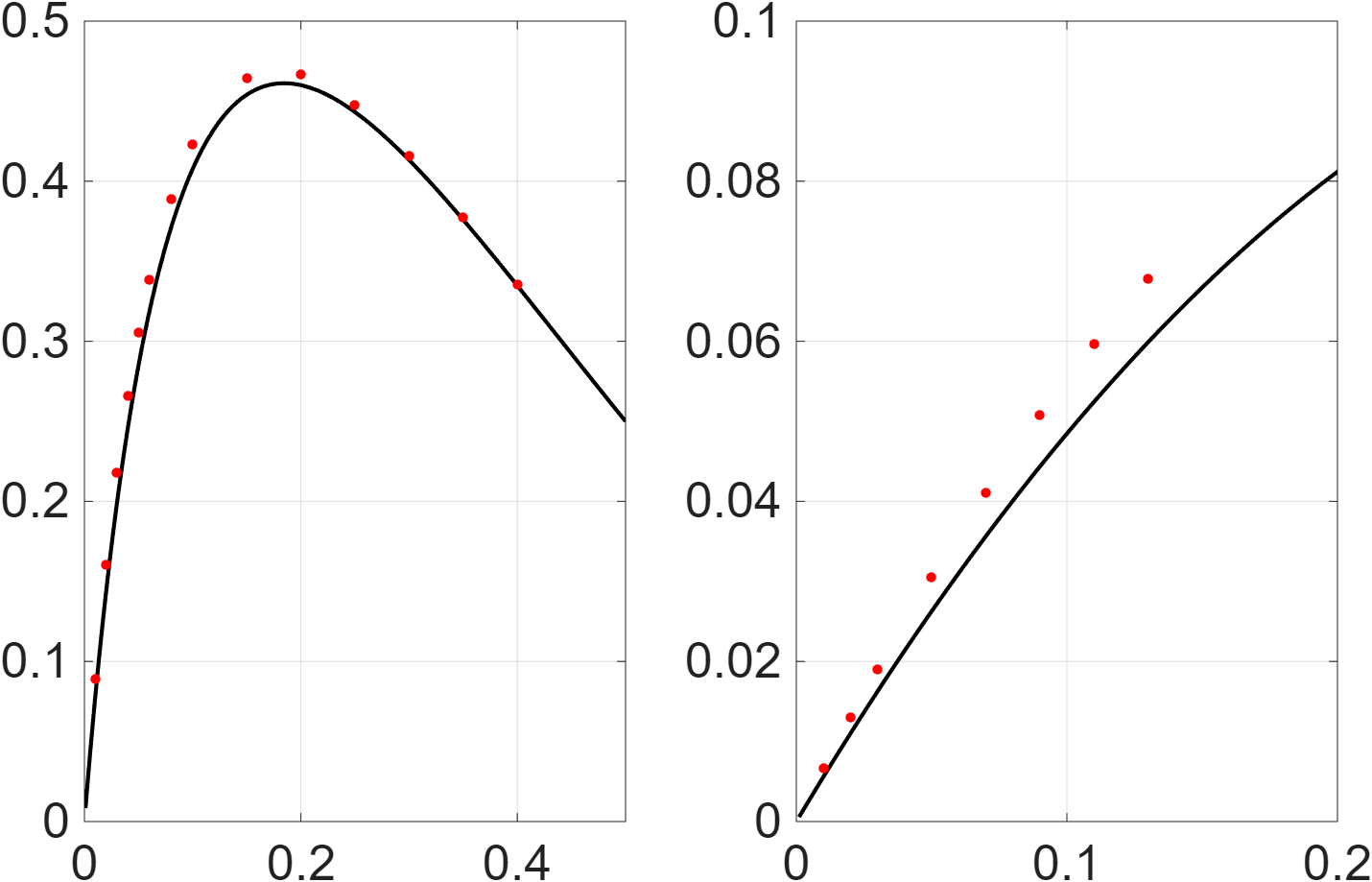}
        };

        \begin{scope}[x={(image.south east)}, y={(image.north west)}]

            \node[above,font=\normalsize] at (.26, .95) {$n_r=1,\ n_\beta=5 $};
            \node[above,font=\normalsize] at (.78, .95) {$n_r=5,\ n_\beta=1 $};

            \node[above,font=\normalsize,black] at (.1, .87) {(a)};
            \node[above,font=\normalsize,black] at (.617, .87) {(b)};
            
            \node[below,font=\normalsize,rotate=0] at (-.06, .55) {$\mathcal{I}^{-1}$};
            \node[below,font=\normalsize] at (0.265, 0) {$\eta$};
            \node[below,font=\normalsize] at (0.78, 0) {$\eta$};

            \draw[fill=white] (.25,-.09) rectangle (.8,-.2);
            
                \node[right,font=\normalsize] at (.32,-.14) {$\mathcal{I}^{-1}$};
                \draw[-,black,thick] (.27,-.14) -- (.32,-.14);

                \node[right,font=\normalsize] at (.66,-.14) {$I_{\text{opt}}^{-1}$};
                \draw[fill=red,red] (.64,-.14) circle (.6mm);
            
        \end{scope}
    \end{tikzpicture}
    \caption{
        \label{fig:optimized_PA}  
        Optimized parametric amplifier receiver (red dots) Compared to the QFI for $n_r=1, n_\beta=5,$ and $n_\beta=1, n_r=5.$ 
    }
\end{figure}
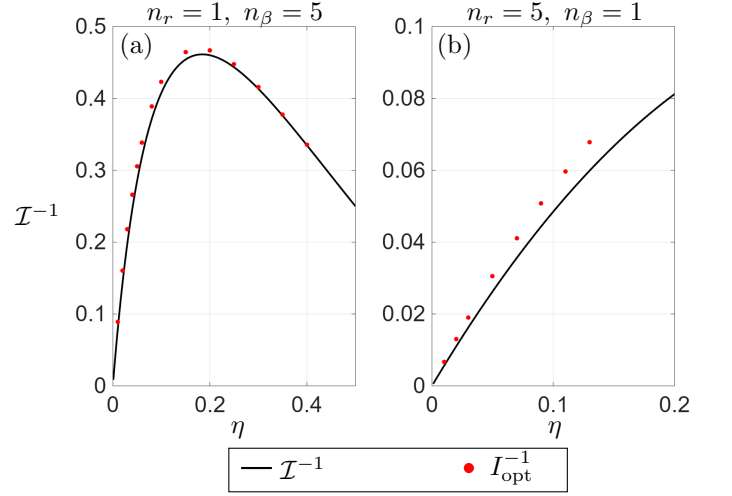

We are now poised to calculate the distribution $p\left(m,n\right)$.
Let us first introduce the Q-function of the state $Q_{\rho}\left(\alpha\right)=\frac{1}{\sqrt{|\pi\Sigma_{Q}|}}\exp\left(-\frac{1}{2}\vec{\alpha}^{\dagger}\Sigma^{-1}_{Q}\vec{\alpha}\right)$, with $\vec{\alpha}=\left(\alpha_{i},\alpha_{s},\alpha^{*}_{i},\alpha^{*}_{s}\right)^{t}$ and $\Sigma_{Q}=\Sigma+\frac{1}{2}I$.
The distribution is then given by \cite{kruse2019detailed}
\begin{align}
p\left(m,n\right)=\frac{1}{n!m!\sqrt{|\Sigma_{Q}|}}\partial^{m}_{\alpha_{i}}\partial^{m}_{\alpha^{*}_{i}}\partial^{n}_{\alpha_{s}}\partial^{n}_{\alpha^{*}_{s}}\exp\left(\frac{1}{2}\vec{\alpha}^{t}\mathcal{A}\vec{\alpha}\right),
\label{eq:probability_from_generating_function}
\end{align}
where $\mathcal{A}=\left(\begin{array}{cc}
0 & \hat{\mathbb{I}}\\
\hat{\mathbb{I}} & 0
\end{array}\right)\left(\hat{\mathbb{I}}-\Sigma^{-1}_{Q}\right)$.
For our state we have
\begin{align}
\begin{split}
&\frac{1}{\sqrt{|\Sigma_{Q}|}}\exp\left(\frac{1}{2}\vec{\alpha}^{t}\mathcal{A}\vec{\alpha}\right)=\\
&\frac{1}{d_{0}}\exp\left[\frac{d_{i}}{d_{0}}|\alpha_{i}|^{2}+\frac{d_{s}}{d_{0}}|\alpha_{s}|^{2}+\frac{C'/2}{d_{0}}\left(\alpha_{i}\alpha_{s}+\alpha^{*}_{i}\alpha^{*}_{s}\right)\right],
\label{eq:generating_function}
\end{split}
\end{align}
where $d_{0}=(1+N_{i})(1+N_{s})-C'^{2}/4,$ $d_{i}=N_{i}(1+N_{s})-C'^{2}/4,$ and $d_{s}=N_{s}(1+N_{i})-C'^{2}/4.$
Inserting Eq.~\ref{eq:generating_function} into Eq.~\ref{eq:probability_from_generating_function} yields the following expression for the distribution
\begin{align}
p\left(m,n\right)=\frac{1}{d^{m+n+1}_{0}}\sum^{\min(m,n)}_{\ell=0}\binom{m}{\ell}\binom{n}{\ell}d^{m-\ell}_{i}d^{n-\ell}_{s}\left(C'/2\right)^{2l}.    
\end{align}
Using this expression we can numerically calculate the FI: $I=\sum_{m,n}\frac{\left(\partial_{\eta}p\left(m,n\right)\right)^{2}}{p\left(m,n\right)}$. $p\left(m,n\right)$ depends on the pre-processing squeezing coefficient $r'$, and we thus numerically optimize this FI with respect to $r'$.
The results are shown in Fig.~\ref{fig:optimized_PA}, where the optimal parametric amplifier FI is compared with the QFI in the squeezing dominated regime of $n_r> n_\eta.$
For $n_r=1, n_\beta=5$ (Fig.~\ref{fig:optimized_PA} (a)) the optimal parametric amplifier FI is very close to the QFI, yet there is still a small gap between them (the largest gap we observed is $7 \%$ for $\eta=0.01$). As expected, this gap closes as $\eta$ approaches the threshold point of $\eta^{*}$. 
For $n_r=5, n_\beta=1$ (Fig.~\ref{fig:optimized_PA} (b)), the gap is larger, where for $\eta=0.01$ the QFI exceeds the corresponding FI by $18\%.$

\section{Non-local Homodyne Optimization}
\label{app:optimization}
\begin{figure}[H]
    \centering
    \begin{tikzpicture}
        \node[anchor=south west, inner sep=0] (image) at (0,0) {
            \includegraphics[width=3.3in, trim={0mm 0mm 0mm 0mm}, clip]{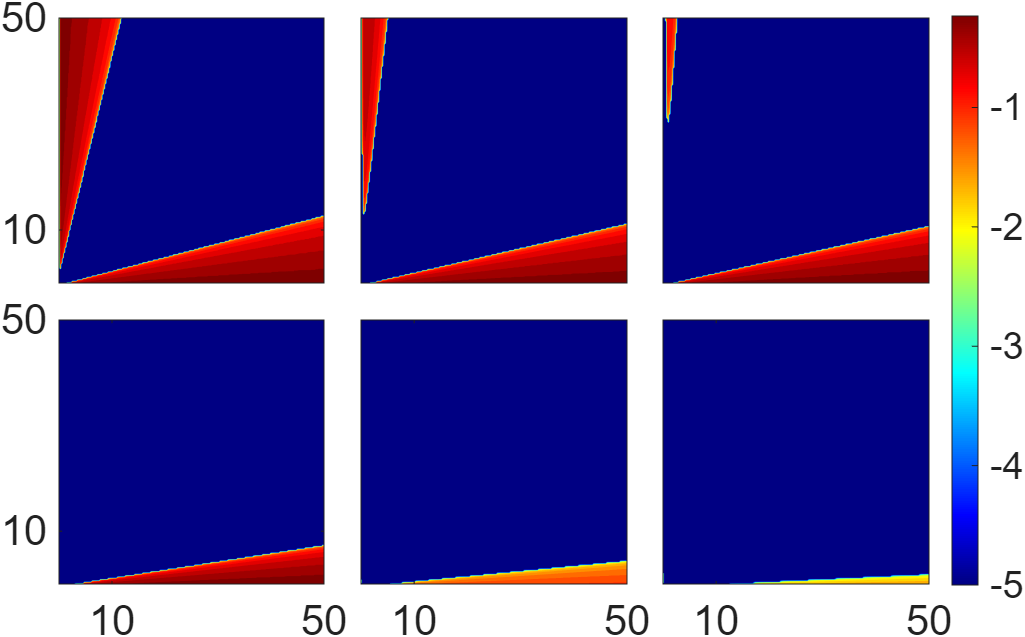}
        };

        \begin{scope}[x={(image.south east)}, y={(image.north west)}]
            \node[below,font=\normalsize] at (-.03, .85) {$n_\beta$};
            \node[below,font=\normalsize] at (-.03, .35) {$n_\beta$};
            \node[below,font=\normalsize] at (1.05, .62) {$r_{opt}$};
            \node[below,font=\normalsize] at (0.2, .02) {$n_r$};
            \node[below,font=\normalsize] at (0.5, .02) {$n_r$};
            \node[below,font=\normalsize] at (0.8, .02) {$n_r$};
            
            \node[below,font=\normalsize,white] at (.23, .98) {$\eta\approx 10^{-3}$};
            \node[below,font=\normalsize,white] at (.54, .98) {$\eta\approx.03$};
            \node[below,font=\normalsize,white] at (.83, .98) {$\eta\approx.05$};
            \node[below,font=\normalsize,white] at (.245, .495) {$\eta\approx.25$};
            \node[below,font=\normalsize,white] at (.54, .495) {$\eta\approx.50$};
            \node[below,font=\normalsize,white] at (.83, .495) {$\eta\approx.75$};

        \end{scope}
    \end{tikzpicture}
    \caption{
        \label{fig:rpost_2d}  
        The optimal pre-processing squeezing parameter $r_\text{opt}$ as a function of $n_r$ and $n_\beta$.
    }
\end{figure}
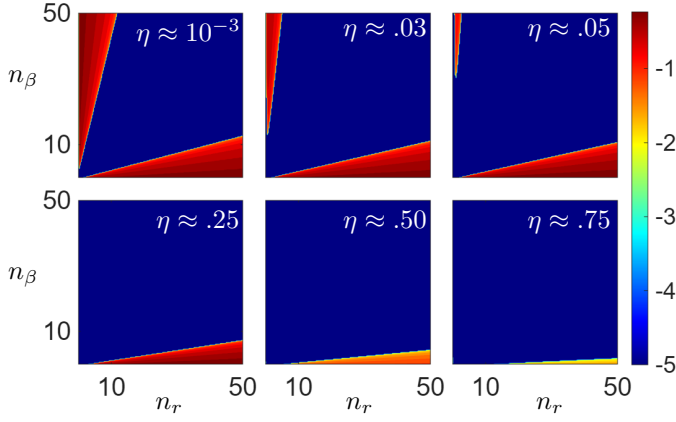

Fig.~\ref{fig:rpost_2d} shows
$r_{\text{opt}}\equiv\underset{r_{\text{pre}}}{\text{argmax }}I_{\text{h,pa}}\left(r_{\text{pre}}\right)$ as a function of $n_r,$ $n_\beta,$ and $\eta.$
For most values of $n_r,n_\beta$, and $\eta,$
the optimum is attained at the lower numerical bound $r_\text{opt}=-5,$ which is numerically equivalent to the limit of $r_\text{pre} \rightarrow \infty.$
This limiting measurement 
corresponds to measuring of $P_{+}, Q_{-}$, hence this measurement is optimal for most of the parameter space.

Deviations from the 
limiting value $r_\text{opt} \rightarrow \infty$
occur primarily in the low reflectivity limit and near the boundaries of the parameter space. In the low-reflectivity regime, $r_\text{opt}$ moves away from the lower bound of $-5,$
becoming less negative, especially in the
signal-dominated ($n_r \gg n_\beta$) and noise-dominated ($n_r \ll n_\beta$) regimes. 
In the signal-dominated regime,
$r_\text{opt}$ exhibits a slow convergence to the lower bound as $\eta$ grows, with deviations persisting until $\eta \to 1$. In contrast, in the noise-dominated regime, $r_\text{opt}$ rapidly converges to 
$-5$ once
$\eta$ exceeds roughly $0.1$. We note that in the intermediate regimes, any $r_\text{opt} \lesssim -2$ already saturates the optimum.
Hence, choosing $r_\text{opt}=-5$ is 
equivalent
to $r_\text{opt}=-\infty$. 
We therefore use $r_\text{opt}=-5$ as the  
lower bound of $r_\text{pre}$
in Fig.~\ref{fig:rpost_2d}.

\section{No entanglement advantage with local homodyne scheme \label{apx:local_homodyne_no_entanglement}}

In this appendix we show that using local homodyne with TMSV is equivalent to using a signal-only scheme of bright single mode squeezed vacuum (SMSV) state and homodyne.
Hence local homodyne scheme cannot outperform the best single-mode scheme and thus cannot offer an entanglement advantage.

The idea is that the homodyne measurement of the idler commutes with the thermal loss channel of the signal, and the protocol is thus equivalent to first performing homodyne of the idler and then sending the signal mode through the thermal encoding channel.
Let us find the state of the signal after homodyne measurement of the idler.
The wavefunction of the TMSV in the $(Q_i,Q_s)$ eigenbasis is
\begin{align}
\psi\left(q_{i},q_{s}\right)=\frac{1}{\sqrt{\pi}}\exp\left[-\frac{\left(q_{i}+q_{s}\right)^{2}}{4e^{2r}}-\frac{\left(q_{i}-q_{s}\right)^{2}}{4e^{-2r}}\right].    
\end{align}
Performing homodyne measurement on the idler mode leads to an outcome x, with a probability of $p\left(x\right)=\frac{1}{\sqrt{\pi\cosh\left(2r\right)}}\exp\left(-x^{2}/\cosh\left(2r\right)\right).$
Given this outcome $x$, the state of the signal mode becomes
\begin{align}
\psi_{s|x}=\left(\frac{\cosh\left(2r\right)}{\pi}\right)^{1/4}\exp\left[-\frac{1}{2}\cosh\left(2r\right)\left(q_{s}-\tanh\left(2r\right)x\right)^{2}\right].    
\end{align}
It is therefore a bright SMSV state with
a mean vector of $\vec{\mu}=\left(\begin{array}{cc}
\tanh\left(2r\right)x & 0\end{array}\right)^{t}$, and a covariance matrix $V=\text{diag}\left\{ \frac{1}{2\cosh\left(2r\right)},2\cosh\left(2r\right)\right\}$.
Local homodyne readout is thus equivalent to a single-mode probabilistic strategy in which this SMSV state is prepared with a probability of $p\left(x\right)=\frac{1}{\sqrt{\pi\cosh\left(2r\right)}}\exp\left(-x^{2}/\cosh\left(2r\right)\right)$.
After passing through the thermal loss channel, this state is given by $\vec{\mu}=\left(\sqrt{\eta}\begin{array}{cc}
\tanh\left(2r\right)x & 0\end{array}\right)^{t}$, 
$V=\text{diag}\left\{ \eta\frac{1}{2\cosh\left(2r\right)}+\sigma^*,2\eta\cosh\left(2r\right)+\sigma^*\right\}$
where $\sigma^*=\left(1-\eta\right)\left(1/2+n_{\beta}\right)$
. Finally, Measuring $Q_{s}$ yields the following FI
\begin{align}
\frac{\tanh\left(2r\right)^{2}x^{2}}{4\eta\sigma^{2}_{Q}}+\frac{1}{2}\frac{\left[\frac{1}{2\cosh\left(2r\right)}-\left(\frac{1}{2}+n_{\beta}\right)\right]^{2}}{\sigma^{2}_{Q}},    
\end{align}
where $\sigma^{2}_{Q}=\eta\frac{1}{2\cosh\left(2r\right)}+\left(1-\eta\right)\left(1/2+n_{\beta}\right)$ is the variance of this measurement.
The total FI is obtained by averaging over $x$.
Given $p\left(x\right)$ we have $\langle x^{2}\rangle=\cosh\left(2r\right)/2$, hence:
\begin{align}
I=\frac{\sinh\left(2r\right)^{2}}{8\eta\sigma^{2}_{Q}\cosh\left(2r\right)}+\frac{1}{2}\frac{\left[\frac{1}{2\cosh\left(2r\right)}-\left(\frac{1}{2}+n_{\beta}\right)\right]^{2}}{\sigma^{4}_{Q}}. 
\label{supp_eq:homodyne_FI_1}
\end{align}
We can now express the FI with $n_r$,
using the relation $\cosh\left(2 r \right)=2n_r+1.$
Note that $\sigma_Q^2=\frac{1}{2n_{r}+1}\left[\frac{1}{2}+\left(1-\eta\right)\tilde{N}\right],$
inserting this in Eq. ~\ref{supp_eq:homodyne_FI_1} and simplifying yields:
\begin{align}
I=\frac{1}{1+2\left(1-\eta\right)\tilde{N}}\left[\frac{n_{r}\left(n_{r}+1\right)}{\eta}+\frac{2\tilde{N}^{2}}{1+2\left(1-\eta\right)\tilde{N}}\right],    
\end{align}
which is indeed equal to $I_\text{h,local}$ (Eq. ~\ref{eq:HFI}).
The local homodyne strategy is thus equivalent to a single-mode strategy of a bright SMSV with a mean vector $\vec{\mu}=\left(\begin{array}{cc}
\sqrt{\frac{2n_{r}\left(n_{r}+1\right)}{2n_{r}+1}} & 0\end{array}\right)^{t},$
and a covariance matrix $V=\text{diag}\left\{ \frac{1}{4n_{r}+2},4n_{r}+2\right\}.$

\section{Adaptive heterodyne-homodyne measurement}

\label{app:adaptive_heterodyne_homodyne}

\textbf{Heterodyne of the signal followed by homodyne of the idler:}
In this protocol we first perform heterodyne measurement of the signal mode and then homodyne measurement of the idler mode along its mean vector.
The covariance matrix of the signal mode is $\left(\eta n_{r}+\left(1-\eta\right)n_{\beta}+1/2\right)\mathbb{I}_{2}$, therefore the probability of detecting $\alpha$  in the heterodyne measurement is $p \left(\alpha \right)=\frac{1}{\pi\left(\eta n_{r}+\left(1-\eta\right)n_{\beta}+1\right)}\exp\left(-\frac{|\alpha|^{2}}{\eta n_{r}+\left(1-\eta\right)n_{\beta}+1}\right)$.
This measurement yields an FI of
\begin{align}
\frac{\left(n_{\beta}-n_{r}\right)^{2}}{\left(1+n_{\beta}\left(1-\eta\right)+n_{r}\eta\right)^{2}}.
\label{supp_eq:heterodyne_signal_FI}
\end{align}
Let us find the state of the idler mode given an outcome of $\alpha$  in the heterodyne measurement. This state can be written as $\text{Tr}_{2}\left[\left(\mathbb{I}\otimes|\alpha\rangle\langle\alpha| \right)\rho(\eta)\right].$
It is more convenient to write this relation with Wigner functions.
The Wigner function of $\rho(\eta)$ is  $W_{IS}=\frac{1}{4\pi^{2}\sqrt{\text{det}\left(\Sigma\right)}}\exp\left(-\frac{1}{2}{\bf q}^{t}\Sigma^{-1}\boldsymbol{q}\right)$,
where $\Sigma$ is the covariance matrix of Eq. ~\ref{eq:Cov}, that can be written using the following blocks of two-by-two matrices: $\Sigma=\left(\begin{array}{cc}
\Sigma_{ii} & \Sigma_{is}\\
\Sigma_{si} & \Sigma_{ss}
\end{array}\right)$.
The Wigner function of $|\alpha\rangle\langle\alpha|$ is $W_{\alpha}=\frac{1}{2\pi\sqrt{\text{det}(\Sigma_{\alpha})}}\exp\left(-\frac{1}{2}\left({\bf q_{s}}-\sqrt{2}\boldsymbol{\alpha}\right)^{t} \Sigma_{\alpha}^{-1} \left(\boldsymbol{q_{s}}-\sqrt{2}\boldsymbol{\alpha}\right)\right)$, where $\boldsymbol{\alpha}=\left(\text{Re}\left(\alpha\right),\text{Im}\left(\alpha\right)\right)^{t}$ and $\Sigma_{\alpha}=\frac{1}{2}\mathbb{I}_2$. The state of the idler mode after heterodyne measurement of the signal mode is thus:
\begin{align}
\text{Tr}_{2}\left[\left(\mathbb{I}\otimes|\alpha\rangle\langle\alpha| \right)\rho(\eta)\right]=\int W_{IS}\left(\boldsymbol{q}\right)W_{\alpha}\left({\bf q_{s}}\right)\,\text{d}q_{s}\text{d}p_{s}.    
\end{align}
Solving this Gaussian integral we obtain that the state of the idler mode after this measurement is Gaussian with a mean vector of $\vec{\mu}=\Sigma_{is}\left(\Sigma_{ss}+\Sigma_{\alpha}\right)^{-1}\sqrt{2}\left(\begin{array}{c}
\text{Re}\left(\alpha\right)\\
-\text{Im}\left(\alpha\right)
\end{array}\right),$ and a covariance matrix of
$V_{i}=\Sigma_{ii}-\Sigma_{is}\left(\Sigma_{ss}+\Sigma_{\alpha}\right)^{-1}\Sigma_{si}.$
Applying this to our case we obtain that the mean vector is 
$\vec{\mu}=\frac{1}{\sqrt{2}}\frac{\sqrt{\eta}\sinh\left(2r\right)}{\eta n_{r}+\left(1-\eta\right)\left(n_{\beta}\right)+1}\left(\text{Re}\left(\alpha\right),-\text{Im}\left(\alpha\right)\right)^{t},$
and the covariance matrix is $V_{i}=\frac{1}{2}\left[\cosh\left(2r\right)-\frac{1}{2}\frac{\eta\sinh\left(2r\right)^{2}}{\eta n_{r}+\left(1-\eta\right)\left(n_{\beta}\right)+1}\right]\mathbb{I}_{2}.$
Performing a homodyne measurement of the signal mode along its mean vector yields the following FI:
\begin{align}
\begin{split}
&\frac{\bigl(1+n_{\beta}+n_{\beta}\eta-n_{r}\eta\bigr)^{2}n_{r}\left(n_{r}+1\right)}{\eta\left(1+n_{\beta}\left(1-\eta\right)+n_{r}\eta\right)^{2}\left(1+n_{r}(2-\eta)+n_{\beta}\left(1+2n_{r}\right)\left(1-\eta\right)\right)}+\\
&\frac{2\left(1+n_{\beta}\right)^{2}n_{r}^{2}\left(n_{r}+1\right)^{2}}{\left(1+n_{\beta}\left(1-\eta\right)+n_{r}\eta\right)^{2}\left(1+n_{r}(2-\eta)+n_{\beta}\left(1+2n_{r}\right)\left(1-\eta\right)\right)^{2}},
\end{split}
\label{supp_eq:homodyne_idler_FI}
\end{align}
where the first term is the information coming from the mean and the second term is the information coming from the covariance.
Combining the FI from the heterodyne measurement (Eq. \ref{supp_eq:heterodyne_signal_FI}) and the FI from the homodyne measurement (Eq. \ref{supp_eq:homodyne_idler_FI}) we obtain that the total FI is:
\begin{align}
\begin{split}
&I=\frac{\left(n_{\beta}-n_{r}\right)^{2}}{\left(1+n_{\beta}\left(1-\eta\right)+n_{r}\eta\right)^{2}}+\\
&\frac{\bigl(1+n_{\beta}+n_{\beta}\eta-n_{r}\eta\bigr)^{2}n_{r}\left(n_{r}+1\right)}{\eta\left(1+n_{\beta}\left(1-\eta\right)+n_{r}\eta\right)^{2}\left(1+n_{r}(2-\eta)+n_{\beta}\left(1+2n_{r}\right)\left(1-\eta\right)\right)}+\\
& \frac{2\left(1+n_{\beta}\right)^{2}n_{r}^{2}\left(n_{r}+1\right)^{2}}{\left(1+n_{\beta}\left(1-\eta\right)+n_{r}\eta\right)^{2}\left(1+n_{r}(2-\eta)+n_{\beta}\left(1+2n_{r}\right)\left(1-\eta\right)\right)^{2}}.
\end{split}    
\end{align}
In the limit of $\eta \rightarrow0$ this FI is simplified to $I\approx\frac{n_{r}\left(n_{r}+1\right)}{\eta\left(1+2n_{r}+n_{\beta}\left(1+2n_{r}\right)\right)},$ which in the quantum illumination limit of $n_\beta \gg 1 \gg n_r$ coincides with the QFI: $I\approx\frac{n_{r}}{\eta n_{\beta}}.$

\textbf{Heterodyne of the idler followed by homodyne of the signal:}
We show that this case is analogous to the classical coherent state case and thus yields the same homodyne FI.
The heterodyne of the idler mode commutes with the thermal channel of the signal mode. This protocol is therefore the same as applying a heterodyne measurement on the idler mode before sending the signal mode through a thermal channel. It will be easier to analyze this analogous protocol. 
The covariance matrix of the idler mode alone is $\left( \frac{1}{2}+n_r \right) \mathbb{I}_2$.
Hence the probability of detecting $\alpha \in \mathbb{C}$ is
\begin{align}
p\left(\alpha\right)=\frac{1}{\pi\left(1+n_r\right)}\exp\left(-\frac{|\alpha|^{2}}{1+n_r}\right).
\label{eq_app:alpha_distribution_heterodyne}
\end{align}
This measurement clearly does not provide any information about $\eta,$ since it is performed before the encoding channel. 
Let us find the state of the signal mode given an outcome of $\alpha$  in the heterodyne measurement.
Note that the input TMSV state can be written in the Fock basis as $|\psi\rangle\propto\underset{k=0}{\overset{\infty}{\sum}}\lambda^{k}|k\rangle_{I}|k\rangle_{S},$
where $\left\{ |k\rangle\right\} _{k}$ represent the Fock states and $\lambda=\sqrt{\frac{n_{r}}{n_{r}+1}}.$
The signal mode state after detecting $\alpha$  in the heterodyne measurement is given by:
\begin{align}
&|\psi\rangle_{S|\alpha}\propto\underset{k=0}{\overset{\infty}{\sum}}\lambda^{k}\langle\alpha|k\rangle|k\rangle_{S}\propto\underset{k=0}{\overset{\infty}{\sum}}\left(\lambda\alpha^{*}\right)^{k}|k\rangle_{S}\\
\end{align}
where $\alpha^{*}$ is the complex conjugate of $\alpha$.
The resulting signal mode state is thus a coherent state: $|\psi\rangle_{S|\alpha}=|\lambda\alpha^{*}\rangle=|\sqrt{\frac{n_{r}}{n_{r}+1}}\alpha^{*}\rangle.$

This protocol is thus equivalent to a probabilistic strategy, in which with a probability of $p(\alpha)$ (Eq.~\ref{eq_app:alpha_distribution_heterodyne})
the signal input state is the coherent state $|\sqrt{\frac{n_{r}}{n_{r}+1}}\alpha^{*}\rangle.$
The homodyne FI of this coherent state is obtained from Eq.~\ref{eq:FI_homodyne_coherent} by replacing $n_r$ with the average number of photons $|\alpha^2|\frac{n_r}{n_r+1}$:
\begin{align}
I=\frac{2n_{\beta}^{2}}{\left[2n_{\beta}\left(1-\eta\right)+1\right]^{2}}+\frac{\frac{n_{r}}{n_{r}+1}|\alpha|^{2}}{\eta\left[2n_{\beta}\left(1-\eta\right)+1\right]}.
\end{align}
The FI of the total protocol is obtained by averaging this FI over the distribution of $\alpha$ (Eq. ~\ref{eq_app:alpha_distribution_heterodyne}).
Since $\int p\left(\alpha\right)|\alpha|^{2}\text{d}\alpha=n_r+1$
we get that the total FI is exactly the coherent state FI of Eq.~\ref{eq:FI_homodyne_coherent}.

\emergencystretch=2em
\bibliography{apssamp}

\end{document}